\documentclass[11pt]{article}
\begin{document}
\def\a{\alpha}\def\b{\beta}\def\g{\gamma}\def\d{\delta}\def\e{\epsilon}
\def\k{\kappa}\def\l{\lambda}\def\L{\Lambda}\def\s{\sigma}\def\S{\Sigma}
\def\Th{\Theta}\def\th{\theta}\def\om{\omega}\def\Om{\Omega}\def\G{\Gamma}
\def\y{\vartheta}\def\m{\mu}\def\n{\nu}
\def\ws{worldsheet}
\def\susy{supersymmetry}
\def\ts{target superspace}
\def\ks{$\k$--symmetry}
\newcommand{\plabel}{\label}
\renewcommand\baselinestretch{1.5}
\newcommand{\nn}{\nonumber\\}\newcommand{\p}[1]{(\ref{#1})}
\renewcommand{\thefootnote}{\arabic{footnote}}

\thispagestyle{empty}

\begin{flushright}
Preprint DFPD 03/TH/25, \\ FTUV/03-0307,  IFIC/03-33
\\
hep-th/0307153
\end{flushright}

\vspace{1.5cm}


\begin{center}
{\LARGE Various Faces of Type IIA Supergravity}

\vspace{1.5cm}

{\bf Igor A. Bandos$^{\ast\dagger}$, Alexei J.
Nurmagambetov$^{\ast}$ and Dmitri P. Sorokin$^{\ast\ddagger}$}
\vspace{1.5cm} \\ {\it
$^{\ast}$ Institute for Theoretical Physics, NSC KIPT}
\\ {\it UA61108,
Kharkov, Ukraine}
\\  {\it
$^{\dagger}$ Departamento de F\'{\i}sica
Te\'orica and IFIC (CSIC-UVEG)},
\\ {\it 46100-Burjassot
(Valencia), Spain }
\\
$^{\ddagger}$ {\it Dipartimento di Fisica, Universit\'a Degli Studi di
Padova}
\\
{\it e INFN, Sezione di Padova, Via F. Marzolo, 8}\\
{\it 35131 Padova, Italia }
\\
ajn@kipt.kharkov.ua
\\
dmitri.sorokin@pd.infn.it
\\
bandos@ific.uv.es

\end{center}


\begin{abstract}
We derive a duality--symmetric action for type IIA $D=10$
supergravity by the Kaluza--Klein dimensional reduction of the
duality--symmetric action for $D=11$ supergravity with the 3--form
and 6--form gauge field. We then double the bosonic fields arising
as a result of the Kaluza--Klein dimensional reduction and add
mass terms to embrace the Romans's version, so that in its final
form the bosonic part of the action contains the dilaton, NS--NS
and RR potentials of the standard type IIA supergravity as well as
their duals, the corresponding duality relations are deduced
directly from the action. We discuss the relation of our approach
to the doubled field formalism by Cremmer, Julia, L\"u and Pope,
complete the extension of this construction to the supersymmetric
case and lift it onto the level of the proper duality-symmetric
action. We also find a new dual formulation of type IIA $D=10$
supergravity in which the NS--NS two--form potential is replaced
with its six--form counterpart. A truncation of this dual model
produces the Chamseddine's version of $N=1$, $D=10$ supergravity.


\end{abstract}

\newpage

\section{Introduction}
A universal, duality--symmetric, formulation of maximal $D=10$ and
$D=11$ supergravities has proved to be useful for the
understanding of many aspects of superstring and M--theory
including their symmetry structure and the dynamics of various
branes constituting an intrinsic part of these theories. It is
well known that superbranes are sources of antisymmetric tensor
fields of supergravity multiplets and that higher dimensional
superbranes couple to the conventional supergravity tensor fields
as well as to their duals (superbrane worldvolume actions
describing this coupling have been constructed in
\cite{F1,M2,dbranes,M5c,M5kp,bns}). To study interactions of such
branes with supergravity backgrounds, and to derive effective
brane actions from corresponding supergravities \cite{sts}, it is
therefore desirable to have a formulation of supergravities in
which the standard and the dual fields enter the action in a
duality--symmetric way. This should be also useful for studying
anomalies in M--theory and in the superstring theories in the
presence of branes
\cite{witten}--\cite{lm}, 
and for the analysis of subtleties in duality relations between
ordinary and chiral forms \cite{witten1}.
 For the discussion of other problems
concerning a consistent description of supergravity--superbrane
systems see e.g. \cite{bbs}--\cite{stelle}.

In \cite{cjlp} a formalism of ``doubled fields'' related by a
twisted self--duality has been developed for describing in a
uniform duality invariant way gauge and internal symmetries of
maximal supergravities, including $D=11$, $D=10$ type IIA and type
IIB supergravities, and their dimensionally reduced versions. An
interesting (super)algebraic structure underlying this
construction has been found. It has allowed one, by introducing a
twisted self--duality condition, to represent the equations of
motion of dual fields as a Maurer--Cartan zero--curvature equation
with the doubled field strengths playing the role of generalized
connection forms. The construction of \cite{cjlp} is on the mass
shell and involves only the bosonic sector of corresponding
supergravities.

The  duality--symmetric (doubled field) action for the complete
$D=11$ supergravity has been constructed in \cite{bbs} and for
type IIB D=10 supergravity in \cite{dls,dlt}. The construction of
these actions is based on the covariant techniques developed in
\cite{pst}. The (twisted) self--duality relation arises in such
formulations as an equation of motion of the corresponding
physical (doubled) fields. However, a duality--symmetric version
of type IIA $D=10$ supergravity is still lacking. A pseudo--action
for doubled Ramond--Ramond fields of type IIA supergravity
considered in \cite{bkorvp} does not produce all the equations of
motion, namely the duality relations between the doubled fields.
So any modification of the theory, such as a nontrivial
self--interaction of fields as in the case of the M5--brane
\cite{M5c,M5kp}, quantum corrections, coupling to other sources
etc. would require an appropriate modification of the duality
relations which can be hard to guess if they are not yielded by a
proper action. Also the democratic formulation of \cite{bkorvp}
did not involve the dualization of the NS--NS two--form potential
which is required for coupling to the NS5--brane.

The aim of this paper is to fill this gap. We obtain the action
for type IIA $D=10$ supergravity by dimensionally reducing the
duality--symmetric formulation \cite{bbs} of $D=11$ supergravity.
We then double the bosonic fields arising as a result of the
Kaluza--Klein compactification and add massive terms to embrace
the Romans's version \cite{romans}, so that in its final form the
proper action contains the mass, the dilaton, NS--NS and RR
potentials (m, $\phi$, $B^{(2)}$, $A^{(1)}$, $A^{(3)}$) of the
conventional type IIA supergravity \cite{gp,cw,hn} as well as
their duals ($A^{(9)}$,$A^{(8)}$, $B^{(6)}$, $A^{(7)}$,
$A^{(5)}$). This allows one to couple the type IIA supergravity to
all Dp--branes, including a topologically massive D2--brane
\cite{tmd0}. It is also the most appropriate for coupling to the
NS5--brane \cite{bns,kurt1} which carries a $B^{(6)}$ charge and
at the same time interacts with the NS--NS field $B^{(2)}$. The
potential $A^{(9)}$ dual to the mass parameter is required for
coupling type IIA supergravity to domain walls, such as a
D8--brane and an O(rientifold)8-brane \cite{polchinski,or,bkorvp},
which have been studied in the context of a higher dimensional and
supersymmetric generalization of the Randall--Sundrum Brane World
scenario and its promotion to String Theory.

We then show how the actions obtained can be rewritten in a simple
``sigma--model'' form which produces the supersymmetrized
group--theoretical formulation of \cite{cjlp}.

As a by--product we also find a new dual formulation of type IIA
$D=10$ supergravity in which the  NS--NS two--form potential is
not present. It is replaced with its six--form counterpart. This
formulation is characterized by an essentially non--polynomial
coupling of the RR one--form potential to the field strengths of
the RR three--form potential and of the NS--NS six--form potential
$B^{(6)}$, and as a consequence, by a highly non--linear $U(1)$
invariance. A truncation of this dual model produces the
Chamseddine's version \cite{chams} of $N=1$, $D=10$ supergravity.
Note also that, a superfield formulation  of dual $N=1$, $D=10$
supergravity with both the dilaton and the NS--NS two--form field
replaced with their eight--form and six--form counterparts was
considered in \cite{km}. This formulation can also be obtained by
an appropriate truncation of the completely duality--symmetric
action for type IIA supergravity considered in this paper.

In what follows, for simplicity, we will focus on the subsector of
ten--dimensional type IIA supergravity which does not involve the
quartic fermion terms. The reason is that as for any supergravity
theory the basic structure of local supersymmetry transformations
and their appropriate modifications in the duality--symmetric PST
approach \cite{pst} can already be deduced at the quadratic level.
In the case of D=11 (duality--symmetric) supergravity recovering
the quartic fermion terms is reached by the supercovariantization
of the action and of local supersymmetry transformations which
leaves however intact the general structure obtained without these
terms. Moreover, the supercovariantization does not change the PST
part of the action and of the local supersymmetry transformations
since this part is already constructed out of the supercovariant
quantities and, thus, implicitly includes quartic fermionic terms.

``Almost the same" happens with the duality--symmetric version of
type IIA supergravity, whose standard formulation was obtained in
\cite{gp,cw,hn} by the dimensional reduction of the
Cremmer--Julia--Scherk $D=11$ supergravity \cite{cjs}. Saying
``almost the same" we mean that recovering the quartic fermion
terms in the standard type IIA supergravity does not only mean the
supercovariantization of the action and of the local supersymmetry
transformations derived in the absence of these terms, but also
requires adding other fermionic terms (see \cite{hn} for the
discussion of this point). However, as in the case of
duality--symmetric $D=11$ supergravity, any modifications of local
supersymmetry transformations due to the PST approach can already
be deduced in the quadratic fermion approximation. Thus the
reconstruction of the quartic fermion terms in the
duality--symmetric type IIA supergravity can be carried out in the
same way as in the usual type IIA supergravity \cite{gp,cw,hn}.

Following the standard Kaluza--Klein way we shall mainly focus on
features of the dimensional reduction and gauge fixing of an
auxiliary scalar field (the PST scalar) appearing in the bosonic
subsector of the duality-symmetric version of D=11 supergravity
\cite{bbs}. This auxiliary field, entering the action in a
non--polynomial way, is assumed to be a singlet under the local
supersymmetry transformations (see \cite{pst,dlt,dlt1,bbs} for
details). This requires the modification of the local
supersymmetry rules for the D=11 gravitino field. After
dimensional reduction this results in the modification of local
supersymmetry transformations of type IIA gravitini and dilatini.
However, as in the case of their eleven--dimensional counterpart,
on the shell of duality relations the local supersymmetry
transformations of the duality--symmetric type IIA supergravity
coincide with that of the standard version.

Since our starting point is the action for duality--symmetric
$D=11$ supergravity, in Section 2 we briefly discuss the structure
of this theory and its relation to the standard $D=11$
supergravity. Section 3 is devoted to the construction of
duality-symmetric type IIA $D=10$ supergravity. In this section we
present different but classically equivalent forms of the action,
give the analysis of symmetry and dynamical properties of the
model, and establish the connection with the standard formulation
of type IIA supergravity.  In Section 4  we complete our
construction by doubling all bosonic fields of the model in a way
similar to the formalism of doubled fields \cite{cjlp,dlt,llps}.
There the completely duality symmetric action for type IIA
supergravity is presented (Sec. 4.1) and the doubled field
sigma--model representation for a duality symmetric supergravity
action is considered (Sec. 4.2). In Section 5 we gauge fix the PST
scalar of the $D=11$ theory in a way to get an action a l\`a Sen
and Schwarz \cite{ss}. As we shall show, the dimensional reduction
of this action, with the PST scalar gauge fixed along the
compactified direction, results in a new dual formulation of type
IIA $D=10$ supergravity which is the $N=2$ generalization of
\cite{chams} and possesses exotic structure of gauge symmetries
and of local supersymmetry realized in a non--linear way.
Alternatively, this formulation can be obtained from the
conventional type IIA supergravity by replacing the NS--NS
two--form $B^{(2)}$ with its dual $B^{(6)}$. In Conclusion we
discuss the results obtained and in Appendices, for reader's
convenience,  we have collected the notation and conventions used
throughout the paper, as well as details concerning dimensional
reduction and useful identities.

\section{Duality--symmetric D=11 supergravity}

The duality--symmetric action for D=11 supergravity proposed in
\cite{bbs} is
\begin{equation}\plabel{cD11}
S=\int_{{\cal M}^{11}}\, \left[\hat{R}^{\hat{ a}_1\hat{
a}_2}\wedge \hat{\Sigma}_{\hat{  a}_1\hat{  a}_2}+{i\over
3!}\hat{\bar\Psi}\wedge{\cal D}[{1\over
2}(\hat{\omega}+\hat{\tilde{\omega}})]\hat{\Psi}\G^{\hat{
a}_1\hat{ a}_2\hat{ a}_3}\wedge\hat{\Sigma}_{\hat{ a}_1\hat{
a}_2\hat{ a}_3}\right]
\end{equation}
$$-\int_{{\cal M}^{11}}\, \left[ {1\over 2}
(\hat{C}^{(7)}+\hat{\ast}\hat{C}^{(4)})\wedge
(\hat{F}^{(4)}+(\hat{F}^{(4)}-\hat{C}^{(4)}))-{1\over
2}\hat{F}^{(4)}\wedge\hat{\ast}\hat{F}^{(4)}+{1\over
3}\hat{A}^{(3)}\wedge\hat{F}^{(4)}\wedge\hat{F}^{(4)}\right]
$$
$$
+\int_{{\cal M}^{11}}\, {1\over 2}i_{\hat{v}}\hat{\cal
F}^{(4)}\wedge\hat{\ast} i_{\hat{v}}\hat{\cal F}^{(4)},
$$
or in a more symmetric form
$$
S=\int_{{\cal M}^{11}}\, \left[\hat{R}^{\hat{ a}_1\hat{
a}_2}\wedge \hat{\Sigma}_{\hat{  a}_1\hat{  a}_2}+{i\over
3!}\hat{\bar\Psi}\wedge{\cal D}[{1\over
2}(\hat{\omega}+\hat{\tilde{\omega}})]\hat{\Psi}\G^{\hat{
a}_1\hat{ a}_2\hat{ a}_3}\wedge\hat{\Sigma}_{\hat{ a}_1\hat{
a}_2\hat{ a}_3} \right.
$$
$$
 \left.-{1\over 2}
(\hat{C}^{(7)}+\hat{\ast}\hat{C}^{(4)})\wedge
(\hat{F}^{(4)}-{1\over 2}\hat{C}^{(4)})-{1\over 2}
(\hat{C}^{(4)}+\hat{\ast}\hat{C}^{(7)})\wedge
(\hat{F}^{(7)}+{1\over 2}\hat{C}^{(7)})\right]
$$
\begin{equation}\plabel{ds11}
+\int_{{\cal M}^{11}}\, [{1\over{4}}\hat F^{(4)}\wedge \hat\ast
\hat F^{(4)} -{1\over 4}\hat F^{(7)}\wedge \hat\ast\hat F^{(7)}
\end{equation}
$$ +{1\over{4}}i_v\hat{\cal F}^{(4)}\wedge  \hat\ast i_v\hat{\cal
F}^{(4)} -{1\over{4}}i_v\hat{\cal F}^{(7)}\wedge \hat\ast
i_v\hat{\cal F}^{(7)}+{1\over 6}\hat F^{(7)}\wedge \hat F^{(4)}].
$$ where $\hat{\ast}$ is the Hodge operator in $D=11$ (the hat is
put to distinguish it from the $D=10$ Hodge $\ast$).

 Modulo the last term the action \p{cD11} is the conventional
$D=11$ supergravity action written in the same notation as in the
original paper \cite{cjs} except for the coefficient (which is one
in our conventions and one quarter in ``supergravity conventions")
in front of the Einstein-Hilbert term, the first term in \p{cD11}
and \p{ds11}, and the coefficient in the definition of the spin
connection (one quarter vs. one half of \cite{cjs}, see below). To
write the Einstein--Hilbert term and the gravitino kinetic term of
the action in the differential form notation it is convenient to
introduce a form dual to the wedge product of the vielbeine
\begin{equation}\plabel{d11S}
\hat{\Sigma}_{\hat{  a}_1\dots\hat{  a}_n}={1\over
(11-n)!}\e_{\hat{  a}_1\hat{  a}_2\dots\hat{
a}_{11}}\hat{E}^{\hat{ a}_{n+1}}\wedge\dots\wedge\hat{E}^{\hat{
a}_{11}}.
\end{equation}
Other building blocks of the action are the covariant derivative
of the gravitino field
\begin{equation}\plabel{Psi1f}
\hat{\Psi}^{\hat{ \a}}=dX^{\hat
m}\hat{\Psi}^{\hat{ \a}}_{\hat m}\; ,
\end{equation}
\begin{equation}\plabel{Dd11}
{\cal D}\hat{\Psi}^{\hat{ \a}}=d\hat{\Psi}^{\hat{
\a}}-{\hat{\omega}^{\hat{  \a}}}_{~~\hat{
\b}}\wedge\hat{\Psi}^{\hat{  \b}}, \qquad {\hat{\omega}^{\hat{
\a}}}_{~~\hat{  \b}}={1\over 4}\hat{\omega}_{\hat{ a}\hat{
b}}{(\G^{\hat{  a}\hat{  b}})^{\hat{ \a}}}_{\hat{  \b}} \; ,
\end{equation}
the bilinear fermionic terms
\begin{equation}\plabel{F4C}
\hat{C}^{(4)}=-{1\over
4}\hat{\bar\Psi}\wedge \hat{\G}^{(2)}\wedge\hat{\Psi},\qquad
\hat{C}^{(7)}={i\over 4}\hat{\bar\Psi}\wedge
\hat{\G}^{(5)}\wedge\hat{\Psi}\; ,
\end{equation}
the supercovariant connection $\hat{\omega}$ determined by
$d\hat{E}^{\hat a}-\hat{E}^{\hat b}\wedge {\hat{\omega}_{\hat
b}}^{~\hat a}={i\over 4}\hat{\bar \Psi}\G^{\hat a}\wedge
\hat{\Psi}$,
\begin{equation}\plabel{sccon}
\hat{\tilde\omega}_{\hat{m}\hat{a}\hat{b}}=\hat{\omega}_{\hat{m}\hat{a}
\hat{b}}+{i\over 8}\hat{\bar\Psi}^{\hat n}
(\G_{\hat{m}\hat{a}\hat{b}\hat{n}\hat{p}})\hat{\Psi}^{\hat p} \; ,
\end{equation}
and the field strength
\begin{equation}\plabel{F4}
\hat{F}^{(4)}=d\hat{A}^{(3)}
\end{equation}
 of the
three--form gauge field $\hat{A}^{(3)}$.

 The last term of \p{cD11} and the corresponding terms in \p{ds11} encode
 the information on duality relations
 between $\hat{A}^{(3)}$ and a six--form gauge field $\hat{A}^{(6)}$,
which can be derived directly from the action \p{cD11}, and
contains the following (anti--)dual combinations of the field
strengths
\begin{eqnarray}\plabel{calF4}
\hat{\cal F}^{(4)} &=& (\hat{F}^{(4)}-\hat{C}^{(4)})
-\hat{\ast}(\hat{ F}^{(7)}+\hat{C}^{(7)})\,,
\\
\plabel{calF7}
\hat{\cal F}^{(7)} &=& \hat{F}^{(7)}+\hat{C}^{(7)}-\hat{\ast}(\hat{
F}^{(4)}-\hat{C}^{(4)})\; =
-\hat{\ast}\hat{\cal F}^{(4)}\; ,
\end{eqnarray}
where
\begin{equation}\plabel{F7}
\hat{F}^{(7)}=d\hat{A}^{(6)}+\hat{A}^{(3)}\wedge\hat{F}^{(4)}.
\end{equation}
 This
part of the actions is constructed with the use of the space--like
unit vector $\hat{v}_{ {\hat m}}$ composed of derivatives of the
PST scalar $a(x)$ \cite{pst} \footnote{The presence in self--dual
and duality--symmetric actions of the auxiliary vector field which
can be leveled at any direction in space with the use of a local
symmetry (see eq. \p{pst2} below) is similar to (and is actually a
manifestation of) the presence of the unobservable Dirac string in
field--theoretical descriptions of monopoles and dyons. }
\begin{equation}\plabel{v}
\hat{v}_{ {\hat m}}={\partial_{ {\hat m}} a\over
{\sqrt{-\partial_{\hat n} a~\hat{g}^{{\hat n}{\hat p}}~
\partial_{\hat p}a}}}
\end{equation}
and $i_{{\hat v}}\hat{\cal F}^{(n)}$ is the inner product of
${\hat v}$ with $\hat{\cal F}^{(n)}$ $(n=4,7)$
\begin{equation}\plabel{ivF4}
i_{{\hat v}}\hat{\cal F}^{(n)}={1\over (n-1)!} dX^{ {\hat
m}_{n-1}}\wedge \cdots \wedge dX^{{ {\hat m}}_1}{\hat v}_{\hat n}
{\hat g}^{\hat{n}\hat{m}}
 \hat{\cal F}^{(n)}_{ {\hat m}
 {\hat m}_1  \cdots {\hat m}_{n-1}}
\equiv {1\over (n-1)!} \hat{E}^{\hat{a}_{n-1}}\wedge \cdots \wedge
\hat{E}^{\hat{a}_1} \; \hat{v}^{\hat{a}} \hat{\cal F}^{(n)}_{
{\hat a}
 {\hat a}_1  \cdots {\hat a}_{n-1}}
\,.
\end{equation}
It is also convenient to introduce the one--form
\begin{equation}\plabel{v1f}
\hat{v}= dX^{\hat{m}} \hat{v}_{ {\hat m}}={1\over
{\sqrt{-\partial_{\hat n} a\, \hat{g}^{{\hat n}{\hat m}}\,
\partial_{\hat m}a}}} \; da \; .
\end{equation}

The action \p{cD11} possesses (by construction)  $D=11$ general
coordinate and local Lorentz invariance and is also invariant
under the following local supersymmetry transformations
\begin{eqnarray}\plabel{ls11}
\d_{\hat{\e}} a &=& 0\; , \qquad
\nonumber \\
\d_{\hat{\e}}\hat{A}^{(3)}&=&{1\over
2}\hat{\bar\e}\hat{\G}^{(2)}\wedge\hat{\Psi}\, ,
\qquad
\d_{\hat{\e}}\hat{A}^{(6)}=-{i\over 2}
\hat{\bar\e}\hat{\G}^{(5)}\wedge\hat{\Psi}+\d\hat{A}^{(3)}
\wedge\hat{A}^{(3)}\, ,
\nonumber \\
\d_{\hat{\e}}
\hat{E}^{ {\hat{a}}}&=&-{i\over 2}\hat{\bar\e} \hat{\G}^{ {\hat
a}}\hat{\Psi}\, ,\qquad
\\
\plabel{lsP} \d_{\hat{\e}}\hat{\Psi} &=& {{\cal
D}}(\hat{\tilde\omega}) \hat{\e}- {i\over 3!4!} \hat{E}^{\hat{ a}}
\, (\hat{\G}_{ {\hat a} {\hat b}_1\dots {\hat b}_4}+8\hat{\G}_{
{\hat b}_1 {\hat b}_2 {\hat b}_3} \hat{g}_{ {\hat b}_4 {\hat a}})
\, \hat{\e} \, ((\hat{F}^{(4)}-\hat{C}^{(4)})^{ {\hat b}_1\dots
{\hat b}_4}-4\hat{v}^{[ {\hat b}_1}{\hat{\cal F}}^{(4) {\hat b}_2
{\hat b}_3 {\hat b}_4] {\hat c}}\hat{v}_{ {\hat c}})
\nonumber \\
 &=&  \left[{\cal
 D}(\hat{\tilde\omega})
-{1\over
3!}\hat{\ast}(\hat{\G}^{(6)}\wedge[(\hat{F}^{(4)}-\hat{C}^{(4)})
-\hat{v}\wedge i_{\hat{v}}\hat{\cal F}^{(4)}])-\right.
\nonumber \\  && \hspace{3cm} -
 \left.
{2i\over
3!}\hat{\ast}(\hat{\G}^{(3)}\wedge\hat{\ast}[(\hat{F}^{(4)}-\hat{C}^{(4)})
-\hat{v}\wedge i_{\hat{v}}\hat{\cal F}^{(4)}])\right]\hat{\e}\; .
\end{eqnarray}
The variations \p{ls11}, \p{lsP} differ from that of \cite{cjs} by
terms with $i_{\hat{v}}\hat{\cal F}^{(4)}$ and also include the
transformation rule for $\hat{A}^{(6)}$ which has been obtained in
\cite{bbs} from the requirement of the supercovariance of
$\hat{F}^{(7)}+\hat{C}^{(7)}$.

To derive other symmetries of the action and equations of motion
for the gauge fields it is convenient to rewrite the part of the
action containing the gauge fields in an equivalent but manifestly
duality-symmetric form with respect to $\hat{ F}^{(7)}$ and
$\hat{F}^{(4)}$
\begin{eqnarray}\plabel{SAd11}
S_{\hat{A}}&=&\int_{{\cal M}^{11}}\, [{1\over 2}{\hat v}\wedge
({\hat{F}}^{(4)}-\hat{C}^{(4)})\wedge i_{\hat v}\hat{\cal F}^{(7)}
-{1\over 2}{\hat v}\wedge ({\hat{F}}^{(7)}+\hat{C}^{(7)})\wedge
i_{\hat v}\hat{\cal F}^{(4)} +{1\over 6} \hat{F}^{(7)}
\wedge {\hat F}^{(4)}\nonumber \\
&& \hspace{3cm} -{1\over 2}\hat{C}^{(4)}\wedge \hat{F}^{(7)} -{1\over
2}\hat{C}^{(7)}\wedge \hat{F}^{(4)}]\; .
\end{eqnarray}

A general variation of \p{SAd11} is
(see Appendix D)
$$ \d S_{\hat{A}}=\int_{{\cal
M}^{11}} [ \d {\hat v}\wedge {\hat v}\wedge (i_{{\hat v}}\hat{\cal
F}^{(4)}\wedge i_{{\hat v}}\hat{\cal F}^{(7)}) +({\hat v}\wedge
i_{{\hat v}}\hat{\cal F}^{(7)})\wedge \d ({\hat {F}}^{(4)}-{\hat
{C}}^{(4)}) +{1\over 2}({\hat { F}}^{(7)}+{\hat {C}}^{(7)})\wedge
\d ({\hat { F}}^{(4)}-{\hat {C}}^{(4)}) $$
\begin{equation}\plabel{SAvar}
+({\hat v}\wedge i_{{\hat v}}\hat{\cal F}^{(4)})\wedge \d ({\hat
{F}}^{(7)}+{\hat {C}}^{(7)}) -{1\over 2}({\hat { F}}^{(4)}-{\hat
{C}}^{(4)})\wedge \d ({\hat { F}}^{(7)}+{\hat {C}}^{(7)})+{1\over
2} \d {\hat A}^{(3)}\wedge {\hat F}^{(4)}\wedge {\hat F}^{(4)}
\end{equation}
$$-{1\over 2}\d(\hat{C}^{(4)}\wedge\hat{F}^{(7)}) -{1\over
2}\d(\hat{C}^{(7)}\wedge\hat{F}^{(4)})],$$ where we have omitted
the total derivative term.

If we are interested only in the variation of the gauge fields, {\it i.e.}
\begin{equation}\plabel{varF}
\d ({\hat {F}}^{(4)}-{\hat {C}}^{(4)})=d(\d {\hat A}^{(3)})\; , \qquad
\d ({\hat {F}}^{(7)}+{\hat {C}}^{(7)})=d(\d {\hat A}^{(6)}-\d{\hat
A}^{(3)}\wedge {\hat A}^{(3)}) +2\d{\hat A}^{(3)}\wedge {\hat
F}^{(4)}\, ,
\end{equation}
and of the PST scalar \p{v}, the general variation \p{SAvar} reduces to
\begin{eqnarray}\plabel{SAvar1}
\d S_A &=& -\int_{{\cal
M}^{11}} \left[(\d {\hat A}^{(6)}-\d{\hat A}^{(3)}\wedge {\hat
A}^{(3)}+{\d a \over \sqrt{-(\hat{\partial a})^2}}
i_{\hat{v}}\hat{\cal F}^{(7)} )\wedge d(\hat{v}\wedge
i_{\hat{v}}\hat{\cal F}^{(4)})\right.
\nonumber \\
&& \left.
-\left(\d {\hat A}^{(3)}+ {\d a \over \sqrt{-(\hat{\partial
a})^2}} i_{\hat{v}}\hat{\cal F}^{(4)} \right) \wedge
(d(\hat{v}\wedge i_{\hat{v}}\hat{\cal F}^{(7)})+2{\hat v}\wedge
i_{\hat{v}}\hat{\cal F}^{(4)}\wedge {\hat F}^{(4)}) \right].
\end{eqnarray}
 From \p{SAvar1} it is easy to see that in addition to the
conventional gauge transformations
\begin{equation}\plabel{gs}
\d{\hat A}^{(3)}=d{\hat \varphi}^{(2)};\qquad \d{\hat
A}^{(6)}=d{\hat \varphi}^{(5)}-{\hat \varphi}^{(2)}\wedge {\hat
F}^{(4)}
\end{equation}
the action \p{cD11} is invariant under a set of `duality related'
transformations \cite{pst} (PST symmetries)
\begin{equation}\plabel{pst1}
\delta_{\hat{\phi}} a=0\; ,\qquad
\delta_{\hat{\phi}}{\hat A}^{(3)}=da\wedge {\hat \phi}^{(2)}\, , \qquad
\delta_{\hat{\phi}} {\hat A}^{(6)}=da\wedge {\hat \phi}^{(5)}+da\wedge {\hat
\phi}^{(2)} \wedge {\hat A}^{(3)}\, ,
\end{equation}
and
\begin{eqnarray}\plabel{pst2}
\delta_{\hat{\Phi}} a={ \Phi}(\hat{x})\; ,\quad
\delta_{\hat{\Phi}} {\hat A}^{(3)}={-{ \Phi}\over
\sqrt{-(\hat{\partial a})^2}} i_{\hat v}\hat{\cal F}^{(4)}\; ,
\quad \delta_{\hat{\Phi}} {\hat A}^{(6)}={-{ \Phi}\over
\sqrt{-(\hat{\partial a})^2}} i_{\hat v}\hat{\cal
F}^{(7)}+\d_{\hat{\Phi}}{\hat A}^{(3)}\wedge {\hat A}^{(3)}\; .
\end{eqnarray}

Equations of motion of ${\hat A}^{(6)}$ and ${\hat A}^{(3)}$ are
\begin{equation}\plabel{eqm}
 d(\hat{v}\wedge i_{\hat{v}}\hat{\cal F}^{(4)})=0\; ,\qquad
d(\hat{v}\wedge i_{\hat{v}}\hat{\cal F}^{(7)})+2{\hat v}\wedge
i_{\hat{v}}\hat{\cal F}^{(4)}\wedge {\hat F}^{(4)}=0\; .
\end{equation}
The general solution to the equation of motion of $\hat{A}^{(6)}$
is \cite{pst}
\begin{equation}\plabel{gsA3}
\hat{v}\wedge i_{\hat v}\hat{\cal F}^{(4)}=da\wedge
d\hat{\xi}^{(2)}.
\end{equation}
Using the symmetry \p{pst1} with
$\hat{\phi}^{(2)}= \hat{\xi}^{(2)}$ one can obtain from \p{gsA3}
\begin{equation}\plabel{A6}
i_{\hat v}\hat{\cal F}^{(4)}=0\; .
\end{equation}
Note that when \p{A6} is satisfied the action \p{cD11} and the
local supersymmetry transformations \p{ls11}, \p{lsP}  coincide with that of
\cite{cjs}.

Then, in the same way the equation of motion of ${\hat A}^{(3)}$ is
reduced to
\begin{equation}\plabel{A3}
i_{\hat v}\hat{\cal F}^{(7)}=0\, .
\end{equation}
Eqs. \p{A6} and \p{A3} together imply the duality relations
between the ${\hat A}^{(3)}$ and ${\hat A}^{(6)}$ field strengths
\begin{eqnarray}\plabel{dc}
\hat{\cal F}^{(4)}=({\hat{F}}^{(4)}-{\hat
{C}}^{(4)})-\hat{\ast}({\hat {F}}^{(7)}+{\hat
{C}}^{(7)})=0\; , \qquad \nonumber
\\
\hat{\cal F}^{(7)}=({\hat{ F}}^{(7)}+{\hat
{C}}^{(7)})-\hat{\ast}({\hat{F}}^{(4)}-{\hat {C}}^{(4)})=0\; .
\qquad
\end{eqnarray}
 Equation of motion of the PST scalar field $a(x)$ is satisfied identically
as a consequence of the equations of motion \p{eqm}. This is the
Noether identity reflecting the second local PST symmetry \p{pst2}
which implies the auxiliary (pure gauge) nature of the PST scalar
$a(x)$. The symmetry  \p{pst2}  allows us to fix, for instance,
the following gauge $\partial_{ {\hat
m}}a(x)=\d^{\underline{11}}_{ {\hat m}}$ which breaks the manifest
$D=11$ Lorentz and general covariance of the theory down to $D=10$
covariance and establishes the connection with non-covariant
approach by  Sen and Schwarz \cite{ss}.

We should also point out that the duality--symmetric version of
$D=11$ supergravity has the following structure
\begin{equation}\plabel{std11ds}
S=S_{EH}+S_{\hat{\Psi}}+S_{\hat{A}},
\end{equation}
where $S_{EH}$ is the Einstein--Hilbert term, $S_{\hat{\Psi}}$ is
the fermion kinetic term and $S_{\hat{A}}$ is a specific term of
the form of eq. \p{SAd11} which contains the information on the
duality relations. In fact, looking ahead, we can claim that
modulo quartic fermion terms which can not be included by the
supercovariantization of the gauge field strengths any
duality-symmetric supergravity can be presented in the form of the
action containing the Einstein--Hilbert term, kinetic terms of the
fermionic fields and a specific construction a l\`a \p{SAd11}.

To conclude this section let us recall that in the conventional
Cremmer--Julia--Scherk formulation of $D=11$ supergravity
\cite{cjs} $$ S_{CJS}=\int \, \left[\hat{R}^{\hat{ a}_1\hat{
a}_2}\wedge \hat{\Sigma}_{\hat{  a}_1\hat{  a}_2}+{i\over
3!}\hat{\bar\Psi}\wedge{\cal D}[{1\over
2}(\hat{\omega}+\hat{\tilde\omega})]\hat{\Psi}\G^{\hat{ a}_1\hat{
a}_2\hat{ a}_3}\wedge\hat{\Sigma}_{\hat{ a}_1\hat{ a}_2\hat{
a}_3}+{1\over
2}\hat{F}^{(4)}\wedge\hat{\ast}\hat{F}^{(4)}\right]$$
\begin{equation}\plabel{cjs}
-\int_{{\cal M}^{11}}\, \left[{1\over 2}
(\hat{C}^{(7)}+\hat{\ast}\hat{C}^{(4)})\wedge
(\hat{F}^{(4)}+(\hat{F}^{(4)}-\hat{C}^{(4)}))+{1\over
3}\hat{A}^{(3)}\wedge\hat{F}^{(4)}\wedge\hat{F}^{(4)}\right]
\end{equation}
the duality conditions arise as a solution of the second order
equation of motion of the field ${\hat A}^{(3)}$ (see, {\it e.g.},
\cite{bbs,cjlp})
\begin{equation}\plabel{A3eq}
d\left(\hat{\ast}({\hat{ F}}^{(4)}-{\hat {C}}^{(4)})-{\hat A}^{(3)}
\wedge {\hat F}^{(4)}-\hat{C}^{(7)}\right)=0 \; .
\end{equation}
 Eq. (\ref{A3eq}) implies that the differential form under the external
 differential is closed,
hence in space--time with trivial topology its solution is an
exact form $d{\hat A}^{(6)}$,
$$
\hat{\ast}({\hat{ F}}^{(4)}-{\hat {C}}^{(4)})-{\hat A}^{(3)}
\wedge {\hat F}^{(4)}-\hat{C}^{(7)} = d{\hat A}^{(6)}\; ,
$$
which is an equivalent representation of eq. \p{dc} ( $\hat{
F}^{(7)}=d{\hat A}^{(6)}+{\hat A}^{(3)}\wedge {\hat F}^{(4)}$, see
eq. \p{F7}).

\section{Duality--symmetric type IIA D=10 supergravity}

To get a duality--symmetric version of type IIA supergravity we
dimensionally reduce the action \p{cD11} a l\`a Kaluza--Klein.

As a first step we separate one spacelike coordinate of the
11-dimensional spacetime,
 \begin{equation}\plabel{X11}
X^{\hat{m}}= (x^m\, , \, X^{\underline{11}}) \; , \qquad
m=0,1,\ldots , 9 \; ,
\end{equation}
and
choose the following ansatz for the vielbein
and gravitino one--forms (see \cite{gp,cw,hn})
\begin{eqnarray}\plabel{vD11ans}
& \hat{E}^{  a}= & e^{{1\over 12}\phi}E^{a} = e^{{1\over 12}\phi
(x) } dx^m E_m^a(x) \; ,  \qquad \nonumber
\\ & \hat{E}^{ 11}=&
e^{-{2\over 3}\phi}(dX^{\underline{11}}+A^{(1)})\; , \qquad
A^{(1)}=dx^mA_m(x)\; ,
\\
\plabel{grD11ans} & \hat{\Psi}\; = & e^{{1\over 24}\phi}\left(\psi+{1\over
12}\G^{(1)}\G^{ {11}}\l\right)-{2\over 3} e^{-{17\over
24}\phi}\, \l \, (dX^{\underline{11}}+A^{(1)})\; ,
\end{eqnarray}
where $\phi(x)$ is the dilaton field,
$A_m(x)$ is the ten-dimensional $U(1)$
gauge field
(which from the point of view of string theory is identified with
Ramond--Ramond one--form potential),
\begin{eqnarray}\plabel{psi1f}
\psi^{\alpha} = dx^m \psi_m^{\alpha}(x) \; ,
\end{eqnarray}
$\psi_m^{\alpha}(x)$ is the ten--dimensional gravitino
and $\lambda(x)$ is a Majorana fermion field which
appear as a result of the Kaluza--Klein splitting (\ref{grD11ans})
of the $D=11$
metric and eleven--dimensional gravitino \p{Psi1f}.

The $D=10$ decomposition of the gauge fields ${\hat A}^{(3)}$ and
${\hat A}^{(6)}$ is
\begin{equation}\plabel{decom}
{\hat A}^{(3)}=A^{(3)}-dX^{\underline{11}}\wedge B^{(2)}, \qquad
{\hat A}^{(6)}=B^{(6)}+dX^{\underline{11}}\wedge A^{(5)},
\end{equation}
where $B^{(2)}$ is the type IIA NS--NS gauge field and $B^{(6)}$
is its dual.

By use of the standard dimensional reduction procedure reviewed in
Appendices B and C we get the following conventional part
\cite{gp,cw,hn} of the type IIA supergravity action (modulo
quartic fermion terms which we shall denote by ${\cal O}(f^4)$)
$$ S_{conven}=\int_{{\cal M}^{10}}\,\left[-R^{{ a}_1{
a}_2}\wedge \Sigma_{{ a}_1{ a}_2}-{i\over 3!}\bar{\psi}\wedge
{\cal D}\psi\wedge \G^{{ a}_1{ a}_2{  a}_3} \Sigma_{{  a}_1{ a}_2{
a}_3} -{i\over 2}\bar{\l}\G^{ a}{\cal D}\l\wedge \Sigma_{ a}
\right] $$ $$ +\int_{{\cal M}^{10}}\,\left[{1\over
2}d\phi\wedge\ast d\phi-(C^{(9)} -\ast C^{(1)})\wedge d\phi\right]
$$ $$ -\int_{{\cal M}^{10}}\,\left[{1\over 2}e^{-{3\over
2}\phi}F^{(2)}\wedge \ast F^{(2)}+(C^{(8)}-e^{-{3\over 2}\phi}\ast
C^{(2)})\wedge F^{(2)}\right] $$ $$ +\int_{{\cal
M}^{10}}\,\left[{1\over 2}e^{\phi}H^{(3)}\wedge\ast H^{(3)}
-(C^{(7)}-e^{\phi}\ast C^{(3)})\wedge H^{(3)}\right] $$ $$
-\int_{{\cal M}^{10}}\,\left[{1\over 2}e^{-{1\over
2}\phi}F^{(4)}\wedge\ast F^{(4)}+(C^{(6)}-e^{-{1\over 2}\phi}\ast
C^{(4)})\wedge F^{(4)}\right] $$
\begin{equation}\plabel{d10st}
+\int_{{\cal M}^{10}}\, B^{(2)}\wedge dA^{(3)}\wedge
dA^{(3)}+{\cal O}(f^4)\,,
\end{equation}
where the field strengths entering the action are defined as
follows\footnote{For reader's convenience we have collected the
definition of all gauge field strengths of type IIA supergravity
and of their dual in Appendix A.}
\begin{equation}\plabel{Fd10}
F^{(2)}=dA^{(1)},\qquad H^{(3)}=dB^{(2)},\qquad
F^{(4)}=dA^{(3)}-H^{(3)}\wedge A^{(1)}
\end{equation}
and
$$
C^{(9)}=-{i\over 4}\bar{\psi}\G^{{ a}_1{ a}_2{
11}}\l\wedge\Sigma_{{  a}_1{  a}_2},\qquad C^{(1)}=-{i\over
2}\bar{\psi}\G^{  11}\l,$$ $$ C^{(8)}={i\over 4\cdot
4!}e^{-{3\over 4}\phi}\bar{\psi}\G^{{  a}_1{ a}_2{ a}_3{ a}_4{
11}}\wedge \psi\wedge \Sigma_{{  a}_1\dots{  a}_4} -{i\over
16}e^{-{3\over 4}\phi}\bar{\psi}\G^{{  a}_1{ a}_2{
a}_3}\l\wedge\Sigma_{{  a}_1\dots{  a}_3}$$$$-{i\over
3}e^{-{3\over 4}\phi}\ast(\bar{\psi}\wedge \G^{(1)}\l)+{15i\over
8\cdot 4!}e^{-{3\over 4}\phi}\bar{\l}\G^{{ a}_1{ a}_2{
11}}\l\wedge\Sigma_{{  a}_1{  a}_2}, $$ $$ C^{(2)}={i\over
4}e^{{3\over 4}\phi}\bar\psi\G^{ 11}\wedge\psi+{i\over
24}e^{{3\over 4}\phi}\bar\psi\wedge \G^{(1)}\l,$$ $$
C^{(3)}={1\over 4}e^{-{1\over 2}\phi}\bar{\psi}\wedge\G^{(1)}\G^{
11}\wedge\psi +{1\over 4}e^{-{1\over
2}\phi}\bar{\psi}\wedge\G^{(2)}\l,
$$
$$
C^{(7)}={i\over 4}e^{{1\over
2}\phi}\bar{\psi}\wedge\G^{(5)}\wedge\psi-{i\over 4}e^{{1\over
2}\phi}\bar{\psi}\wedge\G^{(6)}\G^{ 11}\l,
$$
$$
C^{(4)}=-{1\over 4}e^{{1\over
4}\phi}\bar{\psi}\wedge\G^{(2)}\wedge \psi-{1\over 8}e^{{1\over
4}\phi}\bar{\psi}\wedge\G^{(3)}\G^{  11}\l+{3\over 64}e^{{1\over
4}\phi}\bar{\l}\G^{(4)}\l, $$
\begin{equation}\plabel{d10C}
C^{(6)}={i\over 4}e^{-{1\over 4}\phi}\bar{\psi}\wedge\G^{(4)}\G^{
11}\wedge\psi-{i\over 8}e^{-{1\over
4}\phi}\bar{\psi}\wedge\G^{(5)}\l+{3i\over 64}e^{-{1\over
4}\phi}\bar{\l}\G^{(6)}\G^{  11}\l\,.
\end{equation}
For further use let us note that the four fermionic terms
$C^{(n)}\wedge C^{(10-n)}$, $C^{(1)}\wedge *C^{(1)}$, $e^{-3/2
\phi} C^{(2)}\wedge *C^{(2)}$, $e^{\phi} C^{(3)}\wedge *C^{(3)}$
and $e^{-1/2 \phi} C^{(4)}\wedge *C^{(4)}$ do not contain dilaton
coupling.

In \p{Fd10} and in what follows we define the field strengths of
the NS--NS field $B^{(2)}$ and of its dual $B^{(6)}$ as $H^{(3)}$
and $H^{(7)}$, to distinguish them from the RR field strengths.

The complete action is
\begin{equation}\plabel{c+dsD11}
S= S_{conven} + S_{d.s.}^{(10)}\; ,
\end{equation}
where $ S_{d.s.}^{(10)}$ is obtained by the dimensional reduction
of the last term in Eq. \p{cD11}
\begin{equation}\plabel{dsD11}
S^{(11)}_{d.s.}=
{1\over 2}\int_{{\cal
M}^{11}}\, i_{\hat{v}}\hat{\cal F}^{(4)} \wedge \hat{\ast}
i_{\hat{v}}\hat{\cal F}^{(4)}
=-{1\over 2}\int_{{\cal
M}^{11}}\,\hat{v}\wedge\hat{\cal F}^{(7)}\wedge
i_{\hat{v}}\hat{\cal F}^{(4)}.
\end{equation}
In this section, to reduce the field $\hat{v}$ \p{v} to ten
dimensions we shall assume that it does not depend on the
compactified coordinate, i.e. ${{\partial a}\over{\partial
x^{\underline{11}}}}=0$. This implies that
\begin{equation}\plabel{v10}
\hat{v}=e^{{1\over 12}\phi}v= e^{{1\over 12}\phi (x)} dx^m v_m(x)
\; ,\qquad
i_{\hat{v}}(dX^{11}+A^{(1)})=0, \qquad v_{ { m}}={\partial_{ {m}}
a\over {\sqrt{-\partial_{n} a~{g}^{{ n}{ p}}~
\partial_{ p}a}}}
\end{equation}
where we have used the explicit form of the inverse metric
\begin{equation}\plabel{ginv}
\hat{g}^{ {\hat{m}\hat{n}}}= e^{-{1\over 6} \phi}
\left(\begin{array}{cc} g^{ {mn}} & -A^{ {m}}\\
-A^{ {n}} & -e^{{3\over 2} \phi}+A_lg^{lp}A_p
\end{array}
\right),
\end{equation}
which follows from \p{vD11ans}.

Thus, the reduction of this part of the action results in
\begin{equation}\plabel{dsD10}
S^{(10)}_{d.s.}=\int_{{\cal M}^{10}}[{1\over 2}v\wedge {\cal
H}^{(7)}\wedge i_v{\cal H}^{(3)}+{1\over 2}v\wedge {\cal
F}^{(6)}\wedge i_v{\cal F}^{(4)}],
\end{equation}
where
\begin{eqnarray}\plabel{calF3}
{\cal H}^{(3)}&=&{
H}^{(3)}-C^{(3)}+e^{-\phi}\ast({H}^{(7)}+C^{(7)})\; ,
\nonumber \\
&&  {\cal H}^{(7)}={H}^{(7)}+C^{(7)}+e^{\phi}\ast ({H}^{(3)}-C^{(3)}) =
e^{\phi}  \ast {\cal H}^{(3)}\, , \;
\end{eqnarray}
\begin{eqnarray}\plabel{calF6}
{\cal F}^{(4)}&=&{F}^{ (4)}-C^{(4)}+e^{{1\over 2}\phi} \ast
({F}^{(6)}+C^{(6)}), \nonumber \\
&&  {\cal
F}^{(6)}={F}^{(6)}+C^{(6)}+e^{-{1\over 2}\phi}\ast
({F}^{(4)}-C^{(4)})= e^{-{1\over 2}\phi} \ast {\cal F}^{(4)}
\end{eqnarray}
are intrinsically dual combinations of the field strengths, and
\begin{eqnarray}\plabel{F6H7}
F^{(6)}& = & dA^{(5)}+A^{(3)}\wedge H^{(3)}-B^{(2)}\wedge dA^{(3)}, \quad
\nonumber \\
H^{(7)} & = &dB^{(6)}+A^{(3)} \wedge dA^{(3)}-F^{(6)} \wedge A^{(1)}
\end{eqnarray}
arise upon the dimensional reduction of $\hat{F}^{(7)}$,
\begin{equation}\plabel{hF7dr}
\hat{F}^{(7)}=H^{(7)}+
F^{(6)}\wedge (dX^{{\underline {11}}} + A^{(1)})\, .
\end{equation}
To summarize, we end up with the following action for
duality-symmetric type IIA D=10 supergravity
$$
S=\int_{{\cal
M}^{10}}\,\left[-R^{{  a}_1{  a}_2}\wedge \Sigma_{{ a}_1{
a}_2}-{i\over 3!}\bar{\psi}\wedge {\cal D}\psi\wedge \G^{{ a}_1{
a}_2{  a}_3} \Sigma_{{  a}_1{  a}_2{  a}_3} -{i\over
2}\bar{\l}\G^{  a}{\cal D}\l\wedge \Sigma_{  a} \right]
$$ $$ +\int_{{\cal M}^{10}}\,\left[{1\over 2}d\phi\wedge\ast
d\phi-(C^{(9)} -\ast C^{(1)})\wedge d\phi\right]
$$ $$ -\int_{{\cal M}^{10}}\,\left[{1\over 2}e^{-{3\over
2}\phi}F^{(2)}\wedge \ast F^{(2)}+(C^{(8)}-e^{-{3\over 2}\phi}\ast
C^{(2)})\wedge F^{(2)}\right] $$ $$ +\int_{{\cal
M}^{10}}\,\left[{1\over 2}e^{\phi}H^{(3)}\wedge\ast H^{(3)} -(C^{
(7)}-e^{\phi}\ast C^{(3)})\wedge H^{(3)}\right] $$ $$ -\int_{{\cal
M}^{10}}\,\left[{1\over 2}e^{-{1\over 2}\phi}F^{(4)}\wedge\ast
F^{(4)}+(C^{(6)}-e^{-{1\over 2}\phi}\ast C^{(4)})\wedge
F^{(4)}\right] $$ $$ +\int_{{\cal M}^{10}}\, B^{(2)}\wedge
dA^{(3)}\wedge dA^{(3)}
$$
\begin{equation}\plabel{d10ds}
+\int_{{\cal M}^{10}}\,\left[{1\over 2}v\wedge {\cal
H}^{(7)}\wedge i_v{\cal H}^{(3)}+{1\over 2}v\wedge {\cal
F}^{(6)}\wedge i_v{\cal F}^{(4)}\right] +{\cal O}(f^4).
\end{equation}
and we are ready to discuss its symmetry structure and equations
of motion of the gauge fields which follow from this action.

To this end, as in the case of  duality--symmetric $D=11$
supergravity, it is convenient to rewrite the action as follows
$$
S=\int_{{\cal M}^{10}}\,\left[-R^{{  a}_1{  a}_2}\wedge \Sigma_{{
a}_1{ a}_2}-{i\over 3!}\bar{\psi}\wedge {\cal D}\psi\wedge \G^{{
a}_1{ a}_2{  a}_3} \Sigma_{{  a}_1{  a}_2{ a}_3} -{i\over
2}\bar{\l}\G^{  a}{\cal D}\l\wedge \Sigma_{  a} \right]
$$ $$ +\int_{{\cal M}^{10}}\,\left[{1\over 2}d\phi\wedge\ast
d\phi-(C^{(9)} -\ast C^{(1)})\wedge d\phi\right]
$$ $$ -\int_{{\cal M}^{10}}\,\left[{1\over 2}e^{-{3\over
2}\phi}F^{(2)}\wedge \ast F^{(2)}+(C^{(8)}-e^{-{3\over 2}\phi}\ast
C^{(2)})\wedge F^{(2)}\right] $$
\begin{equation}\plabel{d10ds1}
 -{1\over
2}\int_{{\cal M}^{10}}\, B^{(2)}\wedge dA^{(3)}\wedge
dA^{(3)}+\int_{{\cal M}^{10}}\, {\cal L}^{(10)}_{d.s.}+{\cal
O}(f^4)
\end{equation}
with
\begin{eqnarray}\plabel{ds10}
{\cal L}^{(10)}_{d.s.}&=& {1\over 2} v \wedge ({H}^{(3)}-{C}^{(3)})
\wedge  i_v
{\cal H}^{(7)} -{1\over 2} v \wedge ({F}^{ (4)}-{C}^{(4)})
\wedge i_v {\cal
F}^{(6)}\nonumber\\ &+& {1\over 2} v \wedge ({F}^{(6)}+{C}^{(6)})
\wedge  i_v
{\cal F}^{ (4)} +{1\over 2} v \wedge ({H}^{(7)}+{C}^{(7)})\wedge  i_v {\cal
H}^{(3)}\nonumber\\ & +& {1\over 2}H^{(3)} \wedge  C^{(7)}+{1\over
2}C^{(3)}\wedge H^{(7)}-{1\over 2}C^{(4)}\wedge F^{(6)}-{1\over
2}C^{(6)}\wedge  F^{(4)}\, .
\end{eqnarray}

The variation of \p{ds10} is
\begin{eqnarray}\plabel{Ldsvar}
 \d {\cal L}_{d.s.}^{(10)}&=-[\d v~ v~ i_v{\cal F}^{ (4)} i_v{\cal F}^{(6)}
+v~ i_v{\cal F}^{(6)} \d ({F}^{ (4)}-{C}^{ (4)})-{1\over
2}({F}^{(6)}+{C}^{ (6)}) \d ({F}^{ (4)}-{C}^{ (4)})\nonumber\\
 &-v~ i_v{\cal F}^{(4)}\d ({F}^{(6)} +{C}^{ (6)})
 +{1\over
2}({F}^{ (4)}-{C}^{ (4)})\d ({F}^{(6)}+{C}^{ (6)})+\d v~ v~
i_v{\cal H}^{(7)} i_v{\cal H}^{(3)}\nonumber\\
 &-v~ i_v{\cal
H}^{(7)}\d ({H}^{(3)}-C^{(3)})-{1\over 2} ({H}^{(7)}+{C}^{(7)})\d
({H}^{(3)}-{C}^{(3)}) \nonumber\\ &-v~ i_v{\cal H}^{(3)}\d
({H}^{(7)}+{C}^{ (7)})- {1\over 2}({H}^{(3)}-{C}^{ (3)})\d
({H}^{(7)}+{C}^{ (7)})\nonumber \\& -{1\over 2}\d(H^{(3)}C^{(7)})
-{1\over 2}\d(C^{(3)}H^{(7)})+{1\over 2}\d(C^{(6)}F^{(4)})
+{1\over 2}\d(C^{(4)}F^{(6)}) ]
\end{eqnarray}
modulo a total derivative term. For shortness we omit the wedge
product between the differential forms in (\ref{Ldsvar}) and in
some intermediate formulae below. Since we are interested in the
derivation of symmetries and equations of motion of the gauge
fields we ``freeze" the fermions and deal with the following set
of variations of the field strengths $$ \d {F}^{(2)}=d(\d
A^{(1)}),\quad \d H^{(3)}=d(\d B^{(2)}), \quad \d F^{(4)}=d(\d
A^{(3)})-\d(H^{(3)}\wedge A^{(1)}), $$ $$ \d F^{(6)}= d(\d
A^{\prime (5)})+2\d A^{(3)}\wedge  H^{(3)}-2\d B^{(2)}\wedge
dA^{(3)} \; , $$
\begin{equation}\plabel{listvarF}
\d H^{(7)}=d (\d B^{\prime (6)} )
+2 (\d A^{(3)}+\d B^{(2)} \wedge  A^{(1)}) \wedge F^{(4)}-
\d A^{\prime (5)}\wedge  F^{(2)} - \d A^{(1)}\wedge  F^{(6)} \; ,
\end{equation}
where
\begin{eqnarray}\plabel{varA'}
\d A^{\prime (5)} = \d A^{(5)}+\d B^{(2)} \wedge A^{(3)}-
\d A^{(3)}\wedge  B^{(2)}\; ,
\\ \plabel{varB'}
\d B^{\prime (6)} = \d B^{(6)}-\d A^{(3)}\wedge  A^{(3)}+\d A^{\prime (5)}
\wedge A^{(1)} \; .
\end{eqnarray}

After some algebra one can find that \p{Ldsvar} takes the form

\begin{eqnarray}\plabel{Ldsvar1}
\d {\cal L}^{(10)}_{d.s.}&=&-\left ( {\d a \over \sqrt{-(\partial
a)^2}} i_v{\cal H}^{(3)}+\d B^{(2)} \right )\wedge d(v\wedge i_v{\cal
H}^{(7)})
\nonumber \\  && + \left ( {\d a \over \sqrt{-(\partial a)^2}} i_v{\cal
F}^{(4)}+\d A^{(3)} \right )\wedge d(v \wedge i_v{\cal F}^{(6)})
\nonumber \\  &&
 -\left
( {\d a \over \sqrt{-(\partial a)^2}} i_v{\cal F}^{(6)}+\d
A^{\prime (5)} \right ) \wedge  d(v \wedge i_v{\cal F}^{(4)})
\nonumber \\  &&  -\left ( {\d
a \over \sqrt{-(\partial a)^2}} i_v{\cal H}^{
 (7)}+\d B^{(6)}-\d A^{(3)} \wedge  A^{(3)}+\d
A^{\prime (5)} \wedge  A^{(1)} \right ) \wedge  d(v \wedge i_v{\cal H}^{(3)})
\nonumber \\  &&
+ \d
A^{(1)}   \wedge \left [v \wedge i_v{\cal H}^{(3)}
 \wedge F^{(6)}- v \wedge i_v{\cal F}^{(6)}
\wedge  H^{(3)} +F^{(6)} \wedge  H^{(3)}\right ] \nonumber \\ &&
+\d B^{(2)}  \wedge \left[ d(v  \wedge i_v{\cal F}^{(6)}\wedge
A^{(1)})+2v  \wedge i_v {\cal H}^{(3)}
 \wedge dA^{(3)}  \wedge
A^{(1)}-2v  \wedge i_v {\cal F}^{(4)}  \wedge dA^{(3)}\right]
\nonumber \\ && +\d A^{(3)}  \wedge \left[2v  \wedge i_v{\cal
F}^{(4)}  \wedge H^{(3)} -2v  \wedge i_v{\cal H}^{(3)}\wedge
F^{(4)}\right] \nonumber \\ && +\d A^{\prime (5)}  \wedge v \wedge
i_v{\cal H}^{(3)}
 \wedge F^{(2)}+\d({1\over 2}B^{(2)}  \wedge
dA^{(3)}  \wedge dA^{(3)})\; .
\end{eqnarray}

Analyzing this variation we conclude that in addition to the
conventional gauge symmetries
$$ \d A^{(1)}=d \a^{(0)}\, ,\ \ \ \d B^{(2)}=d \a^{(1)}\, ,\ \ \ \d
A^{(3)}=d \a^{(2)}-B^{(2)}  \wedge d \a^{(0)}\, ,
$$
\begin{equation}\plabel{gs10}
\d A^{(5)}=d \a^{(4)}-d \a^{(2)}  \wedge B^{(2)}+ d \a^{(1)}
 \wedge A^{(3)},\qquad
\d B^{(6)}=d \a^{(5)}-d \a^{(2)}  \wedge A^{(3)}+ d
\a^{(0)}  \wedge A^{(5)}
\end{equation}
the action \p{d10ds1} (or equivalently \p{d10ds}) possesses the
following set of local PST symmetries
$$ \d_{\phi} a(x)=0\, ,\ \ \
\d_{\phi}A^{(1)}=0\, ,\ \ \ \d_{\phi}B^{(2)}=da\wedge \phi^{(1)}\, ,\ \
\ \d_{\phi}A^{(3)}=da\wedge \phi^{(2)}\, , $$
\begin{equation}\plabel{10pst1}
\d_{\phi}A^{\prime (5)}=da\wedge \phi^{(4)}\, ,\qquad
\d_{\phi}B^{(6)}=da\wedge \phi^{(5)}+\d_{\phi}A^{(3)} \wedge A^{(3)}-
\d_{\phi}A^{\prime (5)} \wedge  A^{(1)}\, ,
\end{equation}
 and
$$ \d_{\Phi} a(x)=\Phi(x)\, ,\ \ \ \d_{\Phi} B^{(2)}=-{\Phi \over
\sqrt{-(\partial a)^2}} i_v{\cal H}^{(3)},\ \ \ \d_{\Phi}
A^{(3)}=-{\Phi \over \sqrt{-(\partial a)^2}} i_v{\cal F}^{(4)}\, ,
$$
\begin{equation}\plabel{10pst2}
\d_{\Phi} A^{\prime(5)}=-{\Phi \over \sqrt{-(\partial a)^2}}
i_v{\cal F}^{(6)}\, ,\qquad \d_{\Phi}B^{(6)}=-{\Phi \over
\sqrt{-(\partial a)^2}} i_v{\cal H}^{(7)}+\d_{\Phi}A^{(3)} \wedge
A^{(3)}- \d_{\Phi} A^{\prime (5)} \wedge A^{(1)}\; ,
\end{equation}
where $\d_{\Phi}A^{\prime(5)}$ is defined as in
Eq. \p{varA'}.

Varying the rest of the action \p{d10ds1} and having in mind eq.
\p{Ldsvar1} one gets the following equations of motion of the
gauge fields
\begin{eqnarray}\plabel{Em10}
{\d {\cal L}\over \d B^{(6)}}=0 & \Longrightarrow &
 d(v\wedge i_v{\cal H}^{(3)})=0\, , \nonumber\\
{\d {\cal L}\over \d A^{(5)}}=0 & \Longrightarrow & d(v \wedge i_v
{\cal F}^{(4)})+A^{(1)}\wedge d(v\wedge i_v{\cal H}^{(3)})
-v\wedge i_v{\cal H}^{(3)}\wedge  F^{(2)}=0,\nonumber\\ {\d {\cal
L}\over \d A^{(3)}}=0 &\Longrightarrow & d(v \wedge i_v {\cal
F}^{(6)})+B^{(2)}\wedge  d(v \wedge i_v{\cal F}^{ (4)})
+(A^{(3)}+B^{(2)}\wedge A^{(1)})\wedge d(v\wedge i_v{\cal
H}^{(3)}) \nonumber\\ & &  \hspace{2cm}
 +2v\wedge i_v{\cal
F}^{(4)}\wedge H^{(3)}-2v \wedge i_v{\cal H}^{(3)}\wedge F^{(4)}=0\, ,
\\
 {\d {\cal
L}\over \d B^{(2)}}=0 & \Longrightarrow &
 d(v \wedge i_v{\cal
H}^{(7)})+A^{(3)}\wedge d(v \wedge i_v{\cal F}^{
(4)})+A^{(3)}\wedge  A^{(1)}\wedge d(v \wedge i_v{\cal H}^{(3)})
\nonumber\\ && -
d(v\wedge i_v{\cal F}^{(6)}\wedge  A^{(1)})-2v \wedge i_v{\cal
H}^{(3)}\wedge F^{(4)}\wedge A^{(1)}+2v \wedge i_v{\cal F}^{(4)}\wedge
F^{(4)}=0\, , \nonumber
\end{eqnarray}
\begin{eqnarray}\plabel{Em10'}
{\d {\cal L}\over \d A^{(1)}}=0 & \Longrightarrow &
 d(e^{-{3\over
2}\phi}\ast ({F}^{(2)}-{C}^{(2)})+C^{(8)})+v \wedge i_v{\cal
H}^{(3)} \wedge F^{(6)}
- v \wedge i_v{\cal F}^{(6)} \wedge H^{(3)}\nonumber\\ && \hspace{5cm}
+F^{(6)} \wedge H^{(3)}=0\, ,
\nonumber\\
 {\d {\cal L}\over \d \phi}=0 & \Longrightarrow &
 d[\ast (d\phi-C^{(1)})+C^{(9)}]+{3\over 4}F^{(2)} \wedge
[e^{-{3\over 2}\phi}\ast ({F}^{(2)}-{C}^{(2)})+C^{(8)}]
\nonumber\\ && \hspace{6cm} +{\d {\cal
L}_{d.s.}^{(10)}\over \d \phi}=0\; ,
\end{eqnarray}
where ${\cal L}_{d.s.}^{(10)}$ is defined in \p{ds10}.

Note that these equations are not changed when four-fermion terms
are included (cf. \cite{gp,cw,hn}).

Applying the same arguments as in the case of duality--symmetric
$D=11$ supergravity we can reduce the set of eqs. \p{Em10} to the
duality relations
\begin{equation}\plabel{EmcalF}
{\cal H}^{(3)}=0\; ,\qquad{\cal F}^{(4)}=0\; ,\qquad {\cal F}^{(6)}=0\; ,
\qquad {\cal H}^{(7)}=0\; .
\end{equation}
Then, taking into account \p{EmcalF},  the equations  of motion
\p{Em10'} become
\begin{eqnarray}\plabel{1Em10}
&& d[e^{-{3\over
2}\phi}\ast (F^{(2)}-C^{(2)})+C^{(8)}]-[e^{-{1\over 2}\phi}\ast
({F}^{(4)}-C^{(4)})+C^{(6)}]\wedge  H^{(3)}=0\; , \qquad
\nonumber \\
&& d[\ast (d\phi-C^{(1)})+C^{(9)}]+{3\over 4}F^{(2)}\wedge [e^{-{3\over
2}\phi}\ast (F^{(2)}-C^{(2)})+C^{(8)}]\qquad
\nonumber
\\
&& \qquad + {1\over
2}H^{(3)}\wedge [e^{\phi}\ast ({H}^{(3)}-{C}^{(3)})+C^{(7)}]+{1\over
4}F^{(4)}\wedge [e^{-{1\over 2}\phi} \ast ({F}^{
(4)}-{C}^{(4)})+C^{(6)}]=0\; . \qquad
\end{eqnarray}

Apparently, the equations \p{1Em10} coincide with those obtained
from the action \p{d10st} by varying $A^{(1)}$ and $\phi$. Taking
the external derivative of the duality--symmetric relations
\p{EmcalF} one gets the second order equations of motion of
$B^{(2)}$, $A^{(3)}$ and their duals $B^{(6)}$ and $A^{(5)}$ (see,
for instance, \cite{cjlp}) \footnote{To be precise, in the
standard formulation of type IIA supergravity one can derive the
relations ${\cal F}^{(6)}=0$ and ${\cal H}^{(7)}=0$ by solving
formally the equations of motion for $A^{(3)}$ and $B^{(2)}$. The
relations ${\cal H}^{(3)}=0$ and ${\cal F}^{(4)}=0$ can be
recovered by taking the Hodge dual of the former two.} .
Therefore, we conclude that the duality--symmetric action for type
IIA $D=10$ supergravity is classically equivalent to the
conventional action.

The same observation concerns the local supersymmetry
transformations. By use of the same procedure as in
\cite{gp,cw,hn} one can derive the following supersymmetry
variations of the fields
\begin{eqnarray}\plabel{susy10}
\d_{\e} a& =& 0\, ,\qquad
\nonumber \\
\d_{\e}\phi&=&-{i\over 2}\bar{\e}\G^{  11}\l\; ,
\nonumber \\
 \d_{\e}
A^{(1)}&=&-{i\over 2}e^{{3\over 4}\phi}\bar{\e}\G^{ 11}\psi-{i\over
24}e^{{3\over 4}\phi}\bar{\e}\G^{(1)}\l\; ,
\nonumber \\  \d_{\e}
B^{(2)}&=&{1\over 4}e^{-{1\over 2}\phi}\bar{\e}\G^{(2)}\l+{1\over 2}
e^{-{1\over 2}\phi}\bar{\e}\G^{(1)}\G^{  11}\wedge\psi\; ,
\nonumber \\
\d_{\e} A^{(3)}&=&{1\over 2}e^{{1\over
4}\phi}\bar{\e}\G^{(2)}\psi+{1\over 8} e^{{1\over
4}\phi}\bar{\e}\G^{(3)}\G^{ 11}\l-\d_{\e} B^{(2)}\wedge A^{(1)},
\nonumber \\   \d_{\e} A^{(5)}&=&{i\over 2}e^{-{1\over
4}\phi}\bar{\e}\G^{(4)}\G^{  11}\wedge\psi-{i\over 8}e^{-{1\over
4}\phi}\bar{\e}\G^{(5)}\l+\d_{\e} A^{(3)}\wedge B^{(2)}-\d_{\e}
B^{(2)}\wedge A^{(3)},
\nonumber \\
\d_{\e} B^{(6)}&=&-{i\over 2}e^{{1\over
2}\phi}\bar{\e}\G^{(5)}\wedge\psi+{i\over 4}e^{{1\over
2}\phi}\bar{\e}\G^{(6)}\G^{  11}\l-\d_{\e} A^{(5)}\wedge
A^{(1)}
\nonumber \\
 && + \d_{\e} A^{(3)}\wedge (A^{(3)}+B^{(2)}\wedge A^{(1)})-\d_{\e}
B^{(2)}\wedge A^{(3)}\wedge A^{(1)}\; , \nonumber \\ \d_{\e} E^{
a}&=&-{i\over 2}\bar{\e} \G^{ a}\psi\; , \qquad \nonumber \\
\d_{\e}\psi&=&{\cal D}\e+[{1\over 24}d\phi+{1\over
24}\ast(d\phi\wedge \ast\G^{(2)})+{1\over 4} e^{-{3\over
4}\phi}E^{  a}F^{(2)}_{{  a}{ b}}\G^{ b}\G^{  11} \nonumber \\  &&
+{1\over 3!}\ast(e^{{1\over 2}\phi}\G^{(6)}\wedge[H^{(3)}+v\wedge
i_v{\cal H}^{(3)}]+e^{-{1\over 4}\phi}\G^{(5)}\G^{  11}\wedge[
F^{(4)}-v\wedge i_v{\cal F}^{(4)}]) \nonumber \\   && -{2i\over
3!}\ast(e^{{1\over 2}\phi}\G^{(2)}\G^{
11}\wedge\ast[H^{(3)}+v\wedge i_v{\cal H}^{(3)}]+e^{-{1\over
4}\phi}\G^{(3)}\wedge\ast[F^{(4)}-v\wedge i_v{\cal F}^{(4)}])]\;
\e \nonumber \\ && -{1\over 12}\G^{(1)}\G^{ 11}\d_{\e} \l+{\cal
O}(\e f^2) \; , \nonumber \\ \d_{\e} \l &=& [{1\over 2}\ast(\ast
d\phi\wedge \G^{(1)})\G^{ 11}+{3\over 8}e^{-{3\over
4}\phi}\ast(\ast F^{(2)}\wedge \G^{(2)})+{1\over 4}e^{-{1\over
4}\phi}\ast(\G^{(6)}\wedge [F^{(4)}-v\wedge i_v{\cal F}^{(4)}])
\nonumber \\  && -{i\over 2}e^{{1\over 2}\phi}\ast(\G^{(3)}\wedge
\ast [H^{(3)}+v\wedge i_v{\cal H}^{(3)}])]\; \e+{\cal
O}(\epsilon\, f^2) \; ,
\end{eqnarray}
where ${\cal O}(\epsilon\,f^2)$ stands for terms quadratic in
fermionic fields.

One can see that on the shell of the duality relations \p{EmcalF}
these transformations coincide with that of the standard type IIA
supergravity \cite{gp,cw,hn}.

\section{Completion of the duality--symmetric action}
\subsection{Dualization of the dilaton and of the KK vector field,
and the introduction of the mass term}

One can see that neither the action \p{d10ds} nor \p{d10ds1}
possess the structure of \p{std11ds}. To get such a structure let
us also double the fields $\phi$ and $A^{(1)}$ by introducing
their duals and representing the second order equations of motion
of the former as the Bianchi identities for the dual fields
\cite{cjlp}.

To this end taking into account eqs. \p{EmcalF} one can solve for
eqs. \p{1Em10} in terms of the following dual pairs
\begin{equation}\plabel{DP1}
F^{(1)}=d\phi\,, \quad F^{(9)}=dA^{(8)}-{3\over 4}F^{(8)}\wedge
A^{(1)}+{1\over 2}B^{(2)}\wedge dB^{(6)}-{1\over 4}F^{(6)}\wedge
A^{(3)},
\end{equation}
\begin{equation}\plabel{DP2}
F^{(2)}=dA^{(1)}\; , \qquad F^{(8)}=dA^{(7)}+F^{(6)} \wedge
B^{(2)}+ B^{(2)} \wedge  B^{(2)} \wedge dA^{(3)}\, .
\end{equation}
To include the fermions one should extend the field strengths
\p{DP1} and \p{DP2} as follows
$$ {F}^{(1)}~\rightarrow~{F}^{(1)}-C^{(1)},\qquad
{F}^{(2)}~\rightarrow~{F}^{(2)}-C^{(2)}, $$
\begin{equation}\plabel{tdFd10}
{F}^{(8)}~\rightarrow~{F}^{(8)}+C^{(8)},\qquad
{F}^{(9)}~\rightarrow~{F}^{(9)}+C^{(9)}.
\end{equation}
Then, the following intrinsically dual field strengths
$$ {\cal F}^{(1)}={F}^{(1)}-C^{(1)}+\ast ({F}^{(9)}+C^{(9)}),\qquad
{\cal F}^{(9)}={F}^{(9)}+C^{(9)}+\ast ({F}^{(1)}-C^{(1)})=
\ast {\cal F}^{(1)}, $$
\begin{equation}\plabel{listFadd}
{\cal F}^{(2)}={F}^{(2)}-C^{(2)}+e^{{3\over 2}\phi}\ast
({F}^{(8)}+C^{(8)}),\qquad {\cal
F}^{(8)}={F}^{(8)}+C^{(8)}+e^{-{3\over 2}\phi}\ast
({F}^{(2)}-C^{(2)})= e^{-{3\over 2}\phi} \ast {\cal F}^{(2)}
\end{equation}
are incorporated into the action as follows
\begin{eqnarray}\plabel{compl}
&S=& \int_{{\cal M}^{10}}\, \left(-R^{{  a}_1{  a}_2} \wedge
\Sigma_{{ a}_1{ a}_2}-{i\over 3!}\bar{\psi} \wedge {\cal D}\psi
\G^{{  a}_1{  a}_2{ a}_3} \wedge \Sigma_{{ a}_1{  a}_2{  a}_3}
-{i\over 2}\bar{\l}\G^{ a}{\cal D}\l \wedge \Sigma_{ a}  \right)
\nonumber \\ && +{1\over 2}\sum _{n=1}^{4}\int_{{\cal
M}^{10}}\,\left({1\over{3^{[{{n+1}\over 4}]}}}~ F^{(10-n)}\wedge
F^{(n)}-C^{(10-n)} \wedge  F^{(n)}-F^{(10-n)} \wedge
C^{(n)}\right)
   \\ &&
  + {1\over 2}\sum _{n=1}^{4}\int_{{\cal
M}^{10}} \left[i_v{\cal
F}^{(10-n)}\wedge\,v\wedge({F}^{(n)}-C^{(n)}) +v\wedge
({F}^{(10-n)}+C^{(10-n)})\wedge i_v{\cal F}^{(n)}\right]+{\cal
O}(f^4)\,,\nonumber
\end{eqnarray}
where $[{{n+1}\over 4}]$ denotes the integer part of the number
${{n+1}\over 4}$, and one shall substitute $H^{(3)}$ (${\cal H}^{(3)}$)
and $H^{(7)}$ (${\cal H}^{(7)}$) for $F^{(n)}$ (${\cal F}^{(n)}$)
with $n=3,7$.

This action is the complete duality--symmetric action for type IIA
supergravity up to the four--fermion terms and has the
characteristic structure of duality--symmetric supergravities. The
variation of this action with respect to the gauge fields leads to
the following (PST) gauge fixed set of equations of motion
\begin{equation}\plabel{EMFGF}
{\cal F}^{(1)}={\cal F}^{(2)}={\cal H}^{(3)}={\cal F}^{(4)}={\cal
F}^{(6)}={\cal H}^{(7)}={\cal F}^{(8)}={\cal F}^{(9)}=0\; ,
\end{equation}
which is equivalent to that of the standard formulation.

To extend the action \p{compl} to the Romans's massive
supergravity \cite{romans} let us begin with its bosonic sector in
the form close to that of Ref. \cite{or},
\begin{eqnarray}\plabel{MSG10r}
{\cal L}&=&-R^{ab}\wedge \Sigma_{ab}+{1\over 2}d\phi\wedge
\ast d\phi-{1\over
2}e^{-{1\over 2}\phi}F_m^{(4)}\wedge \ast F_m^{(4)}+{1\over
2}e^{\phi}H_m^{(3)}\wedge \ast H_m^{(3)}
-{1\over 2}e^{-{3\over 2}\phi}F_m^{(2)}\wedge \ast F_m^{(2)}\nonumber
\\ && +dA^{(3)} \wedge
dA^{(3)} \wedge  B^{(2)}+{1\over 3}m\, dA^{(3)}\wedge (B^{(2)})^{\wedge 3}
+{1\over 20}m^2\, (B^{(2)})^{\wedge 5}+
\a m^2 e^{-{5\over 2}\phi}\ast{\bf 1}\; ,
\end{eqnarray}
where $\int_{{\cal M}^{10}}\ast 1 =\int\, d^{10}x \sqrt{-g}$,
$(B^{(2)})^{\wedge 3} \equiv B^{(2)}\wedge B^{(2)}\wedge B^{(2)}$, {\it etc.},
 the parameter $\a$ is to be
fixed by supersymmetry, as in eq. \p{MSGR} below, and
\begin{equation}\plabel{MSGF}
F_m^{(2)}=dA^{(1)}-mB^{(2)}\, ,\quad H_m^{(3)}=H^{(3)}= dB^{(2)}\, ,\quad
F_m^{(4)}=dA^{(3)}-H^{(3)}\wedge A^{(1)}+{1\over 2}m(B^{(2)})^{\wedge 2}\;
\end{equation}
are the `mass--extended' field strengths which are inert under the
 modified gauge transformations
\begin{equation}\plabel{modgt}
\delta B^{(2)}= d\a^{(1)},\quad \delta A^{(1)}=d\a^{(0)}+m
\a^{(1)}\, , \quad \delta A^{(3)}=d\a^{(2)}-B^{(2)}\wedge d\a^{(0)}-m
B^{(2)}\wedge \a^{(1)}\; .
\end{equation}
The action \p{MSG10r}
becomes the bosonic action of the standard `massless' type IIA supergravity
\p{d10st} in the limit $m\rightarrow 0$.

The complete massive type IIA supergravity Lagrangian has the following form
\begin{eqnarray}\plabel{MSGR}
 {\cal L}_{m}&=&{\cal L}_0( F^{(n)}_m)-{9\over
2} e^{-{5\over 2}\phi}m^2 \ast 1 -{3\over 2}e^{-{5\over
4}\phi}m\bar{\l}\G^a\G^{11}\psi\wedge\Sigma_a-{1\over
2}e^{-{5\over 4}\phi}m\bar{\psi}\G^{ab}\wedge\psi\Sigma_{ab} \nonumber \\
&
-& {5\over 4}e^{-{5\over
4}\phi}m \, (\bar{\l}\l)\, \ast 1
+dA^{(3)}\wedge dA^{(3)} \wedge B^{(2)}+{1\over
3}m\, dA^{(3)} \wedge (B^{(2)})^{\wedge 3}+{1\over 20}m^2\,
(B^{(2)})^{\wedge 5}\; . \qquad
\end{eqnarray}
Here we have denoted by ${\cal L}_0(F^{(n)}_m)+dA^{(3)} \wedge
dA^{(3)} \wedge  B^{(2)}$ the Lagrangian for the standard type IIA
supergravity \p{d10st} with the `massless' type IIA supergravity
field strengths $F^{(n)}$ $(n=2,4)$ replaced with $F^{(n)}_m$ of
\p{MSGF}.

The action \p{MSGR} is invariant under modified gauge
transformations \p{modgt} (see e.g. \cite{or}) and local
supersymmetry transformations
$$ \d_{\e}E^{a}=\d_0 E^a\, ,\qquad \d_{\e}\phi=\d_0\phi\,, \qquad
\d_{\e}A^{(n-1)}=\d_0
A^{(n-1)} \quad(n=2,4)\; , \qquad \d_{\e}B^{(2)}=\d_0 B^{(2)}\; ,$$
\begin{equation}\plabel{Msusy10}
\d_{\e}\l=\d_0 \l(F^{(n)}_m)+{3i\over 2}e^{-{5\over
4}\phi}\, m\, \G^{11}\e\; ,\qquad \d_{\e}\psi=\d_0 \psi(F^{(n)}_m)-{i\over
4}e^{-{5\over 4}\phi}\, m\, \G^{(1)}\e\; ,
\end{equation}
where $\d_0$ denotes the `massless' type IIA supersymmetry
transformations, which can be read off \p{susy10}, but with the
appropriate replacement of the field strengths. We note that the
appearance of new terms in the local supersymmetry transformations
is sufficient to completely cancel the contribution from the terms
proportional to the mass parameter $m$ leaving the structure of
the four-fermion terms the same as for the case of massless type
IIA supergravity.

To recover the duality-symmetric structure similar to that of
\p{compl} we double the fields introducing their dual partners by
presenting the second order equations of motion following from
\p{MSGR} as the Bianchi identities \cite{llps}. This procedure
leads to the following set of dual field strenghts for
higher--rank gauge potentials $A^{(5)}$, $B^{(6)}$ and $A^{(7)}$
which are invariant under the modified gauge transformations
\footnote{The modified gauge transformations for $A^{(5)}$,
$B^{(6)}$ and $A^{(7)}$ can be read off the modified field
strengths (\ref{MSGFd0}).}

$$ F^{(6)}_m=F^{(6)}-{1\over 3}m(B^{(2)})^{\wedge 3}\; ,
\qquad H^{(7)}_m=H^{(7)}-m A^{(7)}+{1\over 3}m\, A^{(1)}\wedge
(B^{(2)})^{\wedge 3}\; ,$$
\begin{equation}\plabel{MSGFd0}
F^{(8)}_m=F^{(8)}-{1\over 12}m(B^{(2)})^{\wedge 4}\, , \qquad
F^{(9)}_m=F^{(9)}-{5\over 4}m{A}^{(9)}-{1\over 2}mB^{(2)}\wedge
A^{(7)} +{1\over 16}mA^{(1)}\wedge(B^{(2)})^{\wedge 4}\, ,
\end{equation}
where $F^{(n+5)}$ ($n=1,2,3,4$) are the field strengths defined in
\p{F6H7}, \p{DP1}, \p{DP2}. To derive $F^{(9)}_m$, which can be
read off the dilaton equation of motion
\begin{equation}\plabel{dilEOM}
d \ast d\phi ={1\over 4}F_m^{(4)}\wedge F_m^{(6)}+
{1\over 2}H^{(3)}_m \wedge H^{(7)}_m+
{3\over 4}F^{(2)}_m \wedge F^{(8)}_m-{45\over 4} e^{-{5\over
2}\phi}m^2\, \ast {\bf 1}
\end{equation}
we have used that the right hand side of the above equation is a
ten--form in ten--dimensional space--time and hence is closed and
locally exact. So introducing a nine--form $A^{(9)}$  defined by
\begin{equation}\plabel{A9}
dA^{(9)}=-9e^{-{5\over 2}\phi}m\ast{\bf 1}-B^{(2)}\wedge
F^{(8)}+{1\over 2}(B^{(2)})^{\wedge 2}\wedge F^{(6)}+{1\over
3}dA^{(3)}\wedge (B^{(2)})^{\wedge 3}+{1\over
60}m(B^{(2)})^{\wedge 5}
\end{equation}
one recovers the expression for $F^{(9)}_m$. We should note that
at this stage $A^{(9)}$ is not a dynamical field but an implicit
function of other fields of the theory. It will become a fully
fledged field when the mass parameter is promoted to a scalar
field $F^{(0)}$ dual to a field strength of $A^{(9)}$.

After determining the intrinsically dual combinations of the field
strengths
\begin{eqnarray}
{\cal F}^{(1)}&= & F^{(1)}-C^{(1)}+\ast(F^{(9)}_m+C^{(9)})\;
,\qquad \nonumber \\ && {\cal
F}^{(9)}=F^{(9)}_m+C^{(9)}+\ast(F^{(1)}_m-C^{(1)}) = \ast {\cal
F}^{(1)}\; , \qquad \nonumber \\ {\cal H}^{(3)}&=&
H^{(3)}_m-C^{(3)}+e^{-\phi}\ast(H^{(7)}_m+C^{(7)})\; , \qquad
\nonumber \\ && {\cal
H}^{(7)}=H^{(7)}_m+C^{(7)}+e^{\phi}\ast(H^{(3)}_m-C^{(3)}) =
e^{\phi}\ast {\cal H}^{(3)} \, , \nonumber \\ {\cal F}^{(n)}&=&
F^{(n)}_m-C^{(n)}+e^{{(10-2n)\over
4}\phi}\ast(F^{(10-n)}_m+C^{(10-n)})\; , \quad n=2,4\; , \nonumber
\\  \plabel{MFadd} && {\cal F}^{(n+5)}=
F^{(n+5)}_m+C^{(n+5)}+e^{-{{n}\over
2}\phi}\ast(F^{(5-n)}_m-C^{(5-n)})= e^{-{{n}\over 2}\phi} \ast
{\cal F}^{(5-n)} \; , \quad \nonumber \\ && \hspace{7cm} n=1,3, \;
 \qquad
\end{eqnarray}
the duality-symmetric action for the massive type IIA supergravity
has the following form
\begin{eqnarray}\plabel{Mcompl0}
&S=& \int_{{\cal M}^{10}}\, \left(-R^{{  a}_1{  a}_2} \wedge
\Sigma_{{ a}_1{ a}_2}-{i\over 3!}\bar{\psi} \wedge  {\cal D}\psi
\G^{{  a}_1{  a}_2{ a}_3} \wedge \Sigma_{{ a}_1{  a}_2{  a}_3}
-{i\over 2}\bar{\l}\G^{ a}{\cal D}\l \wedge \Sigma_{ a}  \right)
\nonumber \\ && \quad -\int_{{\cal M}^{10}}\,\left[{9\over
2}e^{-{5\over 2}\phi}m^2\ast {\bf 1}+m(C^{(10)}-e^{-{5\over
2}\phi}\ast C^{(0)})\right] \nonumber \\ &&
\!\!\!\!\!\!\!\!\!\!\!\!\!\!+{1\over 3} \int_{{\cal M}^{10}}\,
m\left( B^{(2)}\wedge F^{(8)}-{1\over 2}(B^{(2)})^{\wedge 2}\wedge
F^{(6)}-{1\over 3}dA^{(3)}\wedge (B^{(2)})^{\wedge 3}-{1\over
60}m(B^{(2)})^{\wedge 5}\right) \nonumber \\ && \quad
\hspace{0.5cm} +{1\over 2}\sum _{n=1}^{4}\int_{{\cal
M}^{10}}\,\left({1\over{3^{[{{n+1}\over 4}]}}}F^{(10-n)}_m\wedge
F^{(n)}_m-C^{(10-n)}\wedge F^{(n)}_m-F^{(10-n)}_m\wedge
C^{(n)}\right)
   \\ &&
  \!\!\!\!\!\!\!\!\!\!\!\!\!\!\!\!\!\!\!\!\!
  + {1\over 2}\sum _{n=1}^{4}\int_{{\cal M}^{10}} \left[i_v{\cal
F}^{(10-n)}\wedge\,v\wedge({F}^{(n)}_m-C^{(n)}) +v\wedge
({F}^{(10-n)}_m+C^{(10-n)})\wedge i_v{\cal F}^{(n)}\right]+{\cal
O}(f^4)\,,\nonumber
\end{eqnarray}
where $[{{n+1}\over 4}]$ denotes the integer part of the number
${{n+1}\over 4}$ and $$ C^{(0)}=-{5\over 4}e^{{5\over
4}\phi}\bar{\l}\l,$$
\begin{equation}\plabel{C10}
C^{(10)}={3\over 2}e^{-{5\over
4}\phi}\bar{\l}\G^a\G^{11}\psi\wedge \Sigma_a+{1\over
2}e^{-{5\over 4}\phi}\bar{\psi}\G^{ab}\wedge\psi\wedge\Sigma_{ab}.
\end{equation}

To extend the action \p{Mcompl0} to a complete duality--symmetric
form we introduce instead of $m$ a zero--form field $F^{(0)}$ and
an exact ten--form $dA^{(9)}$,
and rewrite the Lagrangian \p{MSGR} as follows
$${\cal L}_{m}={\cal
L}_0(F^{(n)}_{m}) -\left[{9\over 2}e^{-{5\over
2}\phi}F^{(0)}\wedge\ast F^{(0)}+(C^{(10)}-e^{-{5\over 2}\phi}\ast
C^{(0)})\wedge F^{(0)}\right]$$
\begin{equation}\plabel{MSGR1}
- F^{(0)}\wedge dA^{(9)}+dA^{(3)}\wedge dA^{(3)}\wedge
B^{(2)}+{1\over 3}F^{(0)}\wedge dA^{(3)}\wedge (B^{(2)})^{\wedge
3}+{1\over 20}(F^{(0)})^2\wedge (B^{(2)})^{\wedge 5}.
\end{equation}
Note that $F^{(0)}$ is inert under the local supersymmetry
transformations, and the equation of motion of $A^{(9)}$,
$dF^{(0)}=0$, implies that $F^{(0)}$ is a constant which we choose
to be $m$. This is an example of the mechanism of the dynamical
generation of mass and tension of branes various aspects of which
have been discussed in the literature \cite{Zh88,tgener}.

 Varying \p{MSGR1} over the new field
$F^{(0)}$ we get the expression for its dual partner $F^{(10)}_m$,
so that together with the dual field strengths \p{MSGFd0} the
complete set becomes
$$ F^{(6)}_m=F^{(6)}-{1\over 3}F^{(0)}(B^{(2)})^{\wedge 3}\; ,
\quad H^{(7)}_m=H^{(7)}-F^{(0)}\wedge A^{(7)}+{1\over 3}F^{(0)}\,
A^{(1)}\wedge (B^{(2)})^{\wedge 3}\; ,$$
$$F^{(8)}_m=F^{(8)}-{1\over 12}F^{(0)}(B^{(2)})^{\wedge 4}\, ,
\quad F^{(9)}_m=F^{(9)}-{5\over 4}F^{(0)}{A}^{(9)}-{1\over
2}F^{(0)}B^{(2)}\wedge A^{(7)} +{1\over
16}F^{(0)}A^{(1)}\wedge(B^{(2)})^{\wedge 4}\, ,$$
\begin{equation}\plabel{MSGFd}
F^{(10)}_m=d{A}^{(9)}+B^{(2)}\wedge F^{(8)}-{1\over
2}(B^{(2)})^{\wedge 2}\wedge F^{(6)}-{1\over 3}(B^{(2)})^{\wedge
3}\wedge dA^{(3)}-{1\over 60}F^{(0)}(B^{(2)})^{\wedge 5}\, .
\end{equation}

After this step one can write the complete duality--symmetric
action for the massive type IIA supergravity as follows
\begin{eqnarray}\plabel{Mcompl}
&S=& \int_{{\cal M}^{10}}\, \left(-R^{{  a}_1{  a}_2} \wedge
\Sigma_{{ a}_1{ a}_2}-{i\over 3!}\bar{\psi} \wedge  {\cal D}\psi
\G^{{  a}_1{  a}_2{ a}_3} \wedge \Sigma_{{ a}_1{  a}_2{  a}_3}
-{i\over 2}\bar{\l}\G^{ a}{\cal D}\l \wedge \Sigma_{ a}  \right)
\nonumber \\ && \quad -\int_{{\cal M}^{10}}\,\left[{9\over
2}e^{-{5\over 2}\phi}F^{(0)}\wedge \ast F^{(0)}
+(C^{(10)}-e^{-{5\over 2}\phi}\ast C^{(0)})\wedge F^{(0)}\right]
\nonumber
\\ && \quad +{1\over 3} \int_{{\cal M}^{10}}\, \left(
F^{(0)}\wedge F^{(10)}_m-dF^{(0)}\wedge A^{(7)}\wedge
B^{(2)}-4F^{(0)}\wedge dA^{(9)}\right) \nonumber
\\ && \quad \hspace{0.5cm} +{1\over 2}\sum _{n=1}^{4}\int_{{\cal
M}^{10}}\,\left({1\over{3^{[{{n+1}\over 4}]}}}F^{(10-n)}_m\wedge
F^{(n)}_m-C^{(10-n)}\wedge F^{(n)}_m-F^{(10-n)}_m\wedge
C^{(n)}\right)
   \\ &&
\!\!\!\!\!\!\!\!\!\!\!\!\!\!\!\!\!\!\!\!\!  + {1\over 2}\sum
_{n=1}^{4}\int_{{\cal M}^{10}} \left[i_v{\cal
F}^{(10-n)}\wedge\,v\wedge({F}^{(n)}_m-C^{(n)}) +v\wedge
({F}^{(10-n)}_m+C^{(10-n)})\wedge i_v{\cal F}^{(n)}\right]+{\cal
O}(f^4)\,,\nonumber
\end{eqnarray}
where as in the eqs. \p{compl}, \p{Mcompl0} $[{{n+1}\over 4}]$
denotes the integer part of the number ${{n+1}\over 4}$. When the
equation $dF^{(0)}=0$ is solved, a constant mass is generated
$F^{(0)}=m$ and the duality relation ${\cal
F}^{(10)}=(F^{(10)}_m+C^{(10)})+9e^{-{5\over 2}\phi}\ast
(F^{(0)}-{1\over 9}C^{(0)})=0$ is taken into account the action
\p{Mcompl} reduces to the action \p{Mcompl0}.

Thus, we have completed our task to construct the
duality--symmetric manifestly covariant version of type IIA
supergravity. We have presented the action in different but
equivalent forms \p{d10ds}, \p{d10ds1} and \p{compl}, \p{Mcompl0},
\p{Mcompl} which serve for different purposes. The action
\p{d10ds} is not manifestly duality--symmetric but its form is
close to the standard action for type IIA supergravity, which
simplifies the verification of the local supersymmetries. The
action \p{d10ds1} is convenient for deriving the duality relations
and for carrying out the symmetry analysis. The actions \p{compl},
\p{Mcompl0} and \p{Mcompl} are manifestly duality--symmetric with
respect to the NS--NS and RR fields and their dual. They can be
considered as an off--shell and supersymmetric generalization of
the democratic formulation of \cite{bkorvp} and of the doubled
field formalism of \cite{cjlp,llps}.

\subsection{The sigma model form of the duality--symmetric supergravity}
Let us discuss the relation of our construction to that of
\cite{cjlp}. In \cite{cjlp} a nice group--theoretical structure
behind the duality relations has been found, which is generic for
all theories in the doubled field formulation. For simplicity, we
shall review this structure with the example of $D=11$
supergravity, however a corresponding sigma--model form of the
duality--symmetric action which we shall present is generic and
valid for all doubled field supergravities.

As it has been noticed in \cite{cjlp}, because of the presence of
the Chern--Simons term the gauge transformations \p{gs} are
non--abelian
\begin{equation}\plabel{gsna}
[\delta_{\Lambda_1^{(3)}},\delta_{\Lambda_2^{(3)}}]=
\delta_{\Lambda^{(6)}}, \quad
[\delta_{\Lambda^{(3)}},\delta_{\Lambda^{(6)}}]=
[\delta_{\Lambda_1^{(6)}},\delta_{\Lambda_2^{(6)}}]=0,
\end{equation}
where $\Lambda^{(3)}$ and $\Lambda^{(6)}$ are closed forms, and
hence they are locally exact $\Lambda^{(3)}=d\hat\varphi^{(2)}$,
$\Lambda^{(6)}=d\hat\varphi^{(5)}$.
 The transformations \p{gsna} can be associated with a superalgebra
generated by a `Grassmann--odd' generator $t_{3}$ and a commuting
(central charge) generator $t_{6}$
\begin{equation}\plabel{salgebra}
\{t_{3},t_{3}\}=-2t_{6}, \qquad [t_{3},t_{6}]=[t_{6},t_{6}]=0.
\end{equation}
The parity of  $t_{3}$ and  $t_{6}$ is related to the
corresponding parity of the differential form potentials
 $\hat A^{(3)}$ and  $\hat A^{(6)}$, so that, for instance $t_{3}$
 anticommutes with $\hat A^{(3)}$ and with the external differential
 $d$.

An element of the supergroup generated by \p{salgebra} can be
realized exponentially as
\begin{equation}\plabel{exp}
{\cal A}=e^{t_{3}\hat A^{(3)}}\,e^{ t_{6}\hat A^{(6)}}\,.
\end{equation}
Then the Cartan form
\begin{equation}\plabel{cf}
{\cal G}=d{\cal A}\,{\cal A}^{-1}=\hat F^{(4)}t_{3} + \hat
F^{(7)}t_{6}
\end{equation}
has the field strengths of the gauge fields $\hat A^{(3)}$ and
$\hat A^{(6)}$ as its components.

By construction, the Cartan form identically satisfies the
Maurer--Cartan equations (the zero curvature condition)
\begin{equation}\plabel{mc}
d{\cal G}+{\cal G}\wedge {\cal G}=0.
\end{equation}
By now in this construction $\hat F^{(4)}$ and $\hat F^{(7)}$ have
been independent field strengths. To impose the duality relation
between them in the framework of this algebraic formalism one
introduces \cite{cjlp} a pseudo--involution ${\cal S}$ which
interchanges the superalgebra generators $t_{3}$ and  $t_{6}$
\begin{equation}\plabel{S}
{\cal S}\,t_{3}=t_{6}, \qquad {\cal S}\,t_{6}=t_{3}, \qquad {\cal
S}^2=1\,.
\end{equation}
 In general,
the eigenvalue of ${\cal S}^2$ on a given generator is the same as
the eigenvalue of $\ast^2$ on the associated field strength. Note
also that the pseudo--involution does not preserve the
superalgebra commutation relations \p{salgebra} \cite{cjlp}.

Using ${\cal S}$ and the Hodge operator one imposes on \p{cf} the
twisted self--duality condition
\begin{equation}\plabel{tsd}
\ast{\cal G}={\cal S}\,{\cal G},
\end{equation}
which reproduces the duality relations between the field strength
components of \p{cf} in the absence of fermions. When \p{tsd}
holds the zero curvature condition \p{mc} amounts to second order
equations of motion of $\hat F^{(4)}$ and $\hat F^{(7)}$.

To add fermions we should extend $\cal G$ with the superalgebra
valued element ${\cal C}=- \hat C^{(4)}t_{3}+{\hat C}^{(7)}t_{6}$
(where $\hat C^{(4)}$ and ${\hat C}^{(7)}$ have been defined in
\p{F4C})
\begin{equation}\plabel{af}
\cal G ~~\rightarrow~~\cal G+\cal C\,.
\end{equation}
Then the twisted self--duality condition takes the form
\begin{equation}\plabel{tsdf}
\ast({\cal G}+{\cal C})={\cal S}\,({\cal G}+{\cal C})\quad
\rightarrow\quad ({\cal S}-\ast)({\cal G}+{\cal C})=0,
\end{equation}
and is tantamount to the duality relations \p{dc}.

We are now ready to present a sigma--model action from which the
twisted self--duality condition \p{tsdf} is obtained as an
equation of motion $$ S=S_{EH}+S_{\hat{\Psi}}-Tr\,\int_{{\cal
M}^{11}}\,{1\over 2}\,({\cal G}+{1\over 2}{\cal C})\wedge({\cal
S}-\ast){\cal C} $$
\begin{equation}\plabel{SAd11sigma}
-Tr\,\int_{{\cal M}^{11}}\,\left\{{1\over 4}\ast{\cal G}\wedge
{\cal G}-{1\over {12}} {\cal G}\wedge {\cal S}{\cal G}  - {1\over
4}\, \ast i_v({\cal S}-\ast)({\cal G}+{\cal C})\wedge i_{v}({\cal
S}-\ast)\,({\cal G}+\cal C) \right\},
\end{equation}
where $S_{EH}$ and $S_{\hat{\Psi}}$ stand for the Einstein action
and the fermion kinetic terms as in \p{std11ds}, the auxiliary one
form $\hat v$ has been introduced in \p{v1f} and the trace is
defined such that
\begin{equation}\plabel{trace}
Tr(t_{3}t_{3})=-Tr(t_{6}t_{6})=-1, \qquad Tr(t_{3}t_{6})=0\,.
\end{equation}
Using the definition of ${\cal G}$ \p{cf} and of the pseudo
involution $\cal S$ \p{S} one can verify that the action
\p{SAd11sigma} is equivalent to the duality--symmetric action
\p{ds11}. One should only note that because of the
anticommutativity of $t_3$ with the odd differential forms the
order of the multipliers in \p{SAd11sigma} is essential.

We should stress that the duality--symmetric action \p{SAd11sigma}
has, actually, a generic form which remains the same also for the
doubled field formulations of type IIA and type IIB $D=10$
supergravities, as well as for lower dimensional supergravities
considered in \cite{cjlp}. To describe a corresponding
supergravity with the action \p{SAd11sigma} one should introduce
the relevant superalgebra \cite{cjlp}, which is analogous but much
more complicated than \p{salgebra}, to construct a corresponding
group element ${\cal A}$ and the Cartan form ${\cal G}$ similar to
\p{exp} and \p{cf}, to define the twisted self--duality condition
\p{tsdf} and insert all these ingredients into the action
\p{SAd11sigma}. We thus have  extended the construction of
\cite{cjlp} to the supersymmetric case and lifting it onto the
level of the proper duality--symmetric action.

In the next section we shall obtain a new version of IIA D=10
supergravity (without the auxiliary PST scalar) which upon the
reduction to N=1 reproduces the six-index photon supergravity by
Chamseddine \cite{chams}.

\section{Gauge fixed version of the duality--symmetric $D=11$ action and a
new exotic formulation of type IIA supergravity}

When we reduced the D=11 supergravity action \p{cD11} to the type
IIA D=10 supergravity action \p{d10ds}, \p{d10ds1} in Section 3 we
chose the PST scalar to be independent of the compactified
coordinate $X^{\underline {11}}$ (eq. \p{v10}). Now we shall
proceed in a different way. Using the symmetry \p{pst2} we
identify the scalar $a(x)$ with the coordinate $X^{\underline
{11}}$. This breaks the D=11 general coordinate invariance of the
duality--symmetric $D=11$ supergravity action down to $D=10$
general coordinate invariance and results in a $D=11$ supergravity
counterpart of the Sen--Schwarz action \cite{ss} for
duality--symmetric gauge fields. Now if we perform the dimensional
reduction of this gauge fixed action along the $X^{\underline
{11}}$ direction, the resulting $D=10$ action does possess the
complete $D=10$ invariance. In addition to the RR fields
$A^{(1)}$, $A^{(3)}$ and the NS--NS field $B^{(2)}$, it also
contains the higher form fields $A^{(5)}$ and $B^{(6)}$ dual to
$A^{(3)}$ and $B^{(2)}$, but it does not involve the PST scalar. A
peculiar feature of this formulation is that the coupling of the
$U(1)$ field $A^{(1)}$ to other fields is non--polynomial and as a
result, the local $U(1)$ symmetry and supersymmetry are realized
in a nonlinear way.

In the gauge $a(x)=X^{\underline {11}}$
 we have $\partial_{ {\hat{m}}}a(x)=
\delta_{ {\hat{m}}}^{\underline {11}}$ and $-\partial
a(x)\hat{g}\partial a(x)= e^{-{1\over 6}\phi}(e^{{3\over 2}\phi}-
A_mg^{mn}A_n)$. Hence
\begin{equation}\plabel{ph2E}
\hat{v}^{(1)}={e^{{1\over 12}\phi} dX^{\underline {11}} \over
\sqrt{e^{{3\over 2}\phi}-(A^{(1)})^2}} \qquad \Leftrightarrow
\qquad \hat{v}_{ {\hat{m}}}=\left( 0\; , {e^{{1\over 12}\phi}
\over \sqrt{e^{{3\over 2}\phi}-(A^{(1)})^2}}\right)\; ,
\end{equation}
{\it i.e.} only $\hat{v}_{\underline {11}}$ component survives,
$\hat{v}_{ {\hat{m}}}= \delta_{ {\hat{m}}}^{\underline
{11}}\hat{v}_{\underline {11}}$. Using \p{ginv}, one can also
check that
\begin{equation}\plabel{ph2E+}
\hat{v}^{ {\hat{m}}}= e^{-{1\over 12}\phi} \left( {A^{ {m}}\over
\sqrt{e^{{3\over 2}\phi}-(A^{(1)})^2}}\; , \; -\sqrt{e^{{3\over
2}\phi}-(A^{(1)})^2}\right)\;
\end{equation}
and
\begin{equation}\plabel{ivG}
i_{\hat v} (dX^{\underline {11}}+A^{(1)})=-{e^{{4\over 3}\phi}
\over \sqrt{ e^{-{1\over 6}\phi}(e^{{3\over 2}\phi}-(A^{(1)})^2})}
\end{equation}
depend solely on the physical fields $\phi(x)$ and $A^{(1)}$,
while $i_{\hat v}\hat{E}^{ {{a}}}=0$ (for $ {{a}}=0, 1, \ldots,
9$). Then the dimensional reduction along the $X^{\underline
{11}}$ direction gives the $D=10$ supergravity action with the
following Lagrangian for the gauge fields
\begin{eqnarray}\plabel{dsLph2G} {\cal
L}^{(10)}_{g.f.}=-{1\over 2}e^{-{3\over 2}\phi}F^{(2)}\wedge \ast
F^{(2)}-(C^{(8)}-e^{-{3\over 2}\phi}\ast C^{(2)})\wedge F^{(2)}
 -{1\over
2}[{\tilde F}^{(4)}\wedge {\cal F}^{(6)}- {\tilde H}^{(7)}\wedge
{\cal H}^{(3)}]
\nonumber \\  +{1\over 2 (A^{(1)2}-e^{{3\over 2}\phi})}[{\tilde
F}^{(4)}\wedge i_A ({\cal H}^{(7)}+{\cal F}^{(6)}A^{(1)})- {\tilde
H}^{(7)}\wedge i_A( {\cal F}^{(4)}+{\cal H}^{(3)}A^{(1)})
]\,\qquad\, \nonumber \\ +{1\over 2}C^{(3)}\wedge
(dB^{(6)}+A^{(3)}\wedge dA^{(3)})-{1\over 2}(C^{(4)}+C^{(3)}\wedge
A^{(1)})\wedge (dA^{(5)}+A^{(3)}\wedge H^{(3)}-B^{(2)}\wedge
dA^{(3)}) \nonumber \\ -{1\over 2}(C^{(7)}+C^{(6)}\wedge
A^{(1)})\wedge H^{(3)}-{1\over 2}C^{(6)}\wedge dA^{(3)}\,,\qquad\,
\end{eqnarray}
where
$$
F^{(2)}=dA^{(1)},\quad \tilde{H}^{(3)}=dB^{(2)}-C^{(3)},\qquad
\tilde{H}^{(7)}=dB^{(6)}+A^{(3)}\wedge dA^{(3)}+C^{(7)}+C^{(6)}\wedge
A^{(1)}\; ,
$$
\begin{equation}\plabel{tilde}
\tilde{F}^{(4)}=dA^{(3)}-C^{(4)}-C^{(3)}\wedge A^{(1)}\; ,
\qquad \tilde{F}^{(6)}=F^{(6)}+C^{(6)},
\end{equation}
and $$ {\cal
H}^{(3)}=\tilde{H}^{(3)}+e^{-\phi}\ast(\tilde{H}^{(7)}
-\tilde{F}^{(6)} \wedge A^{(1)}), \qquad {\cal
H}^{(7)}=\tilde{H}^{(7)}-\tilde{F}^{(6)} \wedge
A^{(1)}+e^{\phi}\ast\tilde{H}^{(3)} =
e^{\phi} * {\cal H}^{(3)}\; ,
$$
\begin{equation}\plabel{Fgf10}
{\cal F}^{(4)}=\tilde{F}^{(4)}-\tilde{H}^{(3)} \wedge A^{(1)}+e^{{1\over
2}\phi}\ast\tilde{F}^{(6)}\; , \qquad {\cal F}^{(6)}=\tilde{F}^{(6)}
+e^{-{1\over 2}\phi}\ast(\tilde{F}^{(4)}-\tilde{H}^{(3)} \wedge A^{(1)})=
e^{-{1\over 2}\phi} *{\cal F}^{(4)}\; .
\end{equation}

Thus, upon the dimensional reduction the gauge fixed $D=11$
supergravity action  reduces to the duality--symmetric type IIA
D=10 supergravity action given by Eq. \p{d10ds1} with ${\cal
L}^{(10)}_{g.f.}$ having the form of \p{dsLph2G}. One could notice
that in \p{dsLph2G} the Kaluza--Klein vector field $A^{(1)}$
couples in a direct non--polynomial way to the field strengths of
$A^{(3)}$ and $B^{(6)}$ and to fermions. However, as we
demonstrate below, the action is nevertheless invariant under
non--manifest local $U(1)$ symmetry associated with $A^{(1)}$.

Note also that in the case under consideration the non--polynomial
structure of $A^{(1)}$ coupling implies the condition
\begin{equation}\plabel{restr}
e^{{3\over 2}\phi}-A_{ {m}} g^{ {m} {n}}A_{ {n}}\ne 0 \;
\end{equation}
which restricts values of the dilaton and of the `length' of the
$U(1)$ gauge field vector. One can also notice that \p{restr} is
the $\hat g^{{\underline {11}},{\underline {11}}}$ component of
the inverse $D=11$ metric \p{ginv}. Actually, since the left hand
side of \p{restr} is not $U(1)$ gauge invariant, this condition
may impose restrictions on the admissible gauge choices for fixing
the $U(1)$ symmetry. However, it is not the case in the Coulomb
gauge in which $A_0=0$, the l.h.s. of \p{restr} is positive
definite (remember that the signature of the metric is
$(+,-,\cdots,-))$ and (when $A_m=0$) tends to zero only in a
non--physical limit $<\phi>\,\rightarrow\,-\infty$. Note also that
\p{restr} is always satisfied in a weak field approximation and
when the $U(1)$ transformations are infinitesimal.

Let us consider what happens with the local symmetries \p{pst1}
and \p{pst2}. When the gauge fixing condition \p{ph2E} is imposed
the transformations \p{pst1} of the $D=11$ action acquire the form
\begin{equation}\plabel{pst1gf}
\d {\hat A}^{(3)}=dX^{ {\underline{11}}}\wedge {\hat
\phi}^{(2)},\qquad \d {\hat A}^{(6)}=dX^{ {\underline {11}}}\wedge
{\hat \phi}^{(5)} +dX^{ {\underline {11}}}\wedge {\hat
\phi}^{(2)}\wedge {\hat A}^{(3)}\; ,
\end{equation}
and thus reduce to local symmetries appearing in the Schwarz--Sen
formulation \cite{ss}. Under the dimensional reduction along the
$X^{\underline {11}}$ direction the gauge potentials are
decomposed as follows
\begin{equation}\plabel{A3dec}
 {\hat A}^{(3)}=A^{(3)}-B^{(2)}\wedge
dX^{\underline {11}}, \qquad {\hat A}^{(6)}=B^{(6)}-A^{(5)}\wedge
dX^{\underline {11}} \; .
\end{equation}
Then the symmetry \p{pst1gf} allows to gauge away the
ten--dimensional gauge field potentials $B^{(2)}$ and $A^{(5)}$.

As far as the PST symmetry \p{pst2} is concerned, although we have
used this symmetry to impose the condition
$a(x)=X^{\underline{11}}$ \p{ph2E}, its combination with the
$U(1)$ gauge transformation (originating in the $D=11$ general
coordinate symmetry and, thus, acting also on
$X^{\underline{11}}$) which preserves \p{ph2E}
\begin{equation}\plabel{KKgfinv}
\d A^{(1)}=d\a^{(0)},\qquad \d X^{\underline {11}}=-\a^{(0)}\,
\qquad \Phi(x)=\a^{(0)}
\end{equation}
is still a local $U(1)$ symmetry of the action (eq. \p{d10gf}
below). Its particular feature is that now also $A^{(3)}$ and
$B^{(6)}$ nontrivially transformed by this $U(1)$: $$ \d
A^{(1)}=d\a^{(0)},$$ $$ \d A^{(3)}=d\a^{(2)}-{\a^{(0)}\over
e^{{3\over 2}\phi}-A^{(1)2}} [i_A{\cal F}^{(4)}+e^{-\phi}\ast
(i_A{\cal H}^{(7)}\wedge A^{(1)})]-\a^{(0)}e^{-\phi}\ast{\cal
H}^{(7)} $$ $$ \doteq d\a^{(2)}-\a^{(0)} e^{-\phi}\ast{\bf
H}^{(7)}\,, $$
\begin{equation}\plabel{gfinv}
\d B^{(6)}=d\a^{(5)}-\delta A^{(3)}\wedge A^{(3)}-{\a^{(0)}\over
e^{{3\over 2}\phi}-A^{(1)2}}(i_A{\cal H}^{(7)}+e^{-{1\over
2}\phi}\ast(i_A{\cal F}^{(4)}\wedge A^{(1)}))-\a^{(0)}e^{-{1\over
2}\phi}\ast {\cal F}^{(4)}
\end{equation}
$$ \doteq d\a^{(5)}-\delta A^{(3)}\wedge
A^{(3)}-\a^{(0)}e^{-{1\over 2}\phi}\ast {\bf F}^{(4)}, $$ where
now, since we have gauged away $B^{(2)}$ and $A^{(5)}$, $${\cal
H}^{(3)}=e^{-\phi}\ast(\tilde{H}^{(7)}-C^{(6)}A^{(1)})-C^{(3)},\qquad
{\cal H}^{(7)}=\tilde{H}^{(7)}-C^{(6)}A^{(1)}-e^{\phi}\ast
C^{(3)},$$
\begin{equation}\plabel{gfcalF}
{\cal F}^{(4)}=\tilde{F}^{(4)}+C^{(3)}A^{(1)}+e^{{1\over
2}\phi}\ast C^{(6)},\qquad {\cal F}^{(6)}=C^{(6)}+e^{-{1\over
2}\phi}\ast(\tilde{F}^{(4)}+C^{(3)}A^{(1)})\; ,
\end{equation}and
\begin{equation}\plabel{iAcalF}
i_A {\cal F}^{(4)}= {1\over 3!} dx^{m_3}\wedge dx^{m_2}\wedge
dx^{m_1}\, A_ng^{mn} {\cal F}^{(4)}_{mm_1m_2m_3}\; , \quad etc.
\end{equation}
and the `boldface' forms
\begin{equation}\plabel{bfH}
{\bf H}^{(7)}={\cal H}^{(7)} +{1\over{e^{{3\over
2}\phi}-A^mA_m}}(e^{\phi}\ast {\cal F}^{(4)}+ i_A {\cal
H}^{(7)})\wedge A^{(1)}\,,
\end{equation}
and
\begin{equation}\plabel{bfF}
{\bf F}^{(4)}=dA^{(3)}+e^{-\phi}\ast{\bf H}^{(7)}\wedge A^{(1)}
\end{equation}
are field strengths which are invariant under the $U(1)$
transformations of \p{gfinv} at least on the mass--shell. For
instance, ${\bf F}^{(4)}$ is invariant only modulo the $B^{(6)}$
field equation of motion
$$
\d_{U(1)}{\bf F}^{(4)}=-\a_0\,d(e^{-\phi}\ast{\bf H}^{(7)}),
$$
where $d(e^{-\phi}\ast{\bf H}^{(7)})=0$ on the mass shell.

As concerns the local supersymmetry transformations, they take the
following form $$ \d_{\e} E^{  a}=-{i\over 2}\bar{\e} \G^{
a}\psi,\qquad \d_{\e}\phi=-{i\over 2}\bar{\e}\G^{ 11}\l, $$ $$
\d_{\e} A^{(1)}=-{i\over 2}e^{{3\over 4}\phi}\bar{\e}\G^{
11}\psi-{i\over 24}e^{{3\over 4}\phi}\bar{\e}\G^{(1)}\l, $$ $$
\d_{\e} A^{(3)}={1\over 2}e^{{1\over
4}\phi}\bar{\e}\G^{(2)}\psi+{1\over 8} e^{{1\over
4}\phi}\bar{\e}\G^{(3)}\G^{  11}\l, $$ $$ \d_{\e} B^{(6)}=-{i\over
2}e^{{1\over 2}\phi}\bar{\e}\G^{(5)}\wedge\psi+{i\over
4}e^{{1\over 2}\phi}\bar{\e}\G^{(6)}\G^{  11}\l+\d_{\e}
A^{(3)}\wedge A^{(3)}, $$ $$ \d_{\e} \l=[{1\over 2}\ast(\ast
d\phi\wedge \G^{(1)})\G^{ 11}+{3\over 8}e^{-{3\over 4}\phi}\ast
(\ast F^{(2)}\wedge \G^{(2)})$$$$ -{1\over 4}e^{{1\over
4}\phi}\ast(\G^{(6)}\wedge \ast [ C^{(6)}-{1\over e^{{3\over
2}\phi}-A^{(1)2}}i_A({\cal H}^{(7)}+{\cal F}^{(6)}\wedge
A^{(1)})-{\cal F}^{(6)}]) $$ $$ +{i\over 2}e^{-{1\over
2}\phi}\ast(\G^{(3)}\wedge[\tilde{H}^{(7)}-C^{(6)}\wedge A^{(1)}+
{1\over e^{{3\over 2}\phi}-A^{(1)2}}i_A({\cal H}^{(7)}+{\cal
F}^{(6)}\wedge A^{(1)}) \wedge A^{(1)}+{\cal F}^{(6)}\wedge
A^{(1)}])]\e+{\cal O}(\epsilon\,f^2) $$ $$ \d_{\e}\psi={\cal
D}\e+[{1\over 24}d\phi+{1\over 24}\ast(d\phi\wedge
\ast\G^{(2)})+{1\over 4} e^{-{3\over 4}\phi}E^{ a}F^{(2)}_{{ a}{
b}}\G^{  b}\G^{  11} $$ $$ -{1\over 3!}\ast(e^{-{1\over
2}\phi}\G^{(6)}\wedge \ast[\tilde{H}^{(7)}-C^{(6)}\wedge A^{(1)}+
{1\over e^{{3\over 2}\phi}-A^{(1)2}}i_A({\cal H}^{(7)}+{\cal
F}^{(6)}\wedge A^{(1)})\wedge A^{(1)}+{\cal F}^{(6)}\wedge
A^{(1)}]$$$$ +e^{{1\over 4}\phi}\G^{(5)}\G^{ 11}\wedge
\ast[C^{(6)}-{1\over e^{{3\over 2}\phi}-A^{(1)2}}i_A({\cal
H}^{(7)}+{\cal F}^{(6)}\wedge A^{(1)})-{\cal F}^{(6)}]) $$ $$
+{2i\over 3!}\ast(e^{{1\over 4}\phi}\G^{(3)}\wedge [C^{(6)}-
{1\over e^{{3\over 2}\phi}-A^{(1)2}}i_A({\cal H}^{(7)}+{\cal
F}^{(6)}\wedge A^{(1)})-{\cal F}^{(6)}]$$
\begin{equation}\plabel{susy10gf}
+e^{-{1\over 2}\phi}\G^{(2)}\G^{
11}\wedge[\tilde{H}^{(7)}-C^{(6)}\wedge A^{(1)}+{1\over e^{{3\over
2}\phi}-A^{(1)2}}i_A({\cal H}^{(7)}+{\cal F}^{(6)}\wedge
A^{(1)})\wedge A^{(1)}+{\cal F}^{(6)}\wedge A^{(1)}])]\e
\end{equation}
$$-{1\over 12}\G^{(1)}\G^{ 11}\d_{\e} \l+{\cal O}(\epsilon\,f^2) $$ with the
field strengths defined in \p{tilde} and \p{gfcalF}.

Consequently, we end up with the following type IIA supergravity
action
$$
S=\int_{{\cal M}^{10}}\,\left[-R^{{  a}_1{  a}_2}\wedge \Sigma_{{
a}_1{ a}_2}-{i\over 3!}\bar{\psi}\wedge {\cal D}\psi\wedge \G^{{
a}_1{ a}_2{  a}_3} \Sigma_{{  a}_1{  a}_2{  a}_3} -{i\over
2}\bar{\l}\G^{  a}{\cal D}\l\wedge \Sigma_{  a} \right]
$$ $$ +\int_{{\cal M}^{10}}\,\left[{1\over 2}d\phi\wedge\ast
d\phi-(C^{(9)} -\ast C^{(1)})\wedge d\phi\right]
$$
\begin{equation}\plabel{d10gf} -\int_{{\cal
M}^{10}}\,\left[{1\over 2}e^{-{3\over 2}\phi}F^{(2)}\wedge \ast
F^{(2)}+F^{(2)}\wedge(C^{(8)}-e^{-{3\over 2}\phi}\ast C^{(2)})
\right]
\end{equation}
$$ -\int_{{\cal M}^{10}}\,\left[{1\over 2}e^{-{1\over 2}\phi} {\bf
F}^{(4)}\wedge \ast{\bf F}^{(4)}+{\bf F}^{(4)}\wedge
(C^{(6)}-e^{-{1\over 2}\phi}\ast (C^{(4)}+{1\over 2}C^{(3)}\wedge
A^{(1)}))\right ] $$ $$ -\int_{{\cal M}^{10}}\,\left[{1\over 2}
e^{-\phi}{\bf H}^{(7)}\wedge \ast {\bf H}^{(7)}+e^{-\phi}\ast {\bf
H}^{(7)}\wedge {\cal H}^{(7)}+{1\over 2}C^{(3)}\wedge {\bf
H}^{(7)}\right]$$$$+\int_{{\cal M}^{10}}\, {1\over 2}C^{(3)}\wedge
(dB^{(6)}+A^{(3)}\wedge dA^{(3)})
 +{\cal O}(f^4)\,,
$$ where, as always in this paper, ${\cal O}(f^4)$ stands for
quartic fermion terms, and $$ {\cal
H}^{(7)}=dB^{(6)}+A^{(3)}\wedge dA^{(3)}+C^{(7)}-e^{\phi}\ast
C^{(3)}\,. $$

We have thus seen that in this new version the $U(1)$ gauge field
potential couples in a non--polynomial way to other fields and, as
a consequence, the gauge symmetries and the local supersymmetry
are realized in a highly non--linear fashion. In the way in which
\p{d10gf} has been obtained, this is the consequence of the
mixture of space--time and PST symmetries caused by their gauge
fixing in the self--dual $D=11$ supergravity which gives rise to
the action \p{d10gf} upon dimensional reduction. However, as we
shall argue below, the nature of this phenomenon is not in a
particular method of dualization but in the presence of the
Maxwell potential $A^{(1)}$ in the field strength $F^{(4)}$
\p{Fd10} of the conventional type IIA supergravity.

 One may wonder how the action \p{d10gf}
is related to the type IIA supergravity in the form \p{d10ds} and
to the conventional action \p{d10st}. Firstly, since \p{d10gf}
does not contain the dual RR field $A^{(5)}$ and its field
strength $F^{(6)}$, to relate \p{d10gf} to \p{d10ds} we should get
rid of $F^{(6)}$ also in the latter. Note that this is easy to do
since the ``bare'' potential $A^{(5)}$ never appears in \p{d10ds}.
Hence, we can use a {\sl non--dynamical} relation $i_v{\cal
F}^{(4)}=0$, which is part of the duality--symmetric equations of
motion \p{EmcalF}, to completely eliminate $F^{(6)}$ from the
action \p{d10ds}.

Secondly, in \p{d10ds} one should also eliminate $B^{(2)}$ and
$H^{(3)}$ replacing them with their dual $B^{(6)}$ and $H^{(7)}$.
One can, modulo a total derivative, rewrite the Chern--Simons term
in \p{d10ds} such that it will contain $H^{(3)}$ instead of
$B^{(2)}$ $$ \int_{{\cal M}^{10}}\, B^{(2)}\wedge dA^{(3)}\wedge
dA^{(3)} \quad \rightarrow \quad \int_{{\cal M}^{10}}\,
H^{(3)}\wedge A^{(3)}\wedge dA^{(3)}\,. $$ Once this is done and
when $F^{(6)}$
 is eliminated, the
action does not contain the ``bare'' potential $B^{(2)}$ anymore,
and its $U(1)$ invariant field strength $H^{(3)}=dB^{(2)}$ can be
replaced with the $U(1)$ invariant field strength ${\bf H}^{(7)}$
\p{bfH} by solving the first of the duality relations \p{EmcalF},
which now reduces to \footnote{Actually, the validity of the
duality relation \p{h3} is an implicit proof of the $U(1)$
invariance of ${\bf H}^{(7)}$ \p{bfH}.}
\begin{eqnarray}\plabel{h3}
H^{(3)}=-e^{ \phi}\,\ast {\bf H}^{(7)}
\end{eqnarray}

Substituting \p{h3} into \p{d10ds} we get the action \p{d10gf}
obtained by an alternative gauge fixing and the dimensional
reduction of the duality--symmetric $D=11$ supergravity.

The same result can also be achieved by a direct dualization of
the field strength $H^{(3)}$ in the conventional type IIA
supergravity action \p{d10st}. For this one should regard
$H^{(3)}$ as an independent field, add to the action \p{d10st} the
Lagrange multiplier term $(H^{(3)}-dB^{(2)})\wedge H^{(7)}_0$ and
replace $H^{(3)}$ with ${\bf H}^{(7)}$ \p{bfH} by solving the
equations of motion for $H^{(3)}$. Note that the equation of
motion of $B^{(2)}$ implies that $H^{(7)}_0=dB^{(6)}$.

So, we should stress that the non--polynomial nature of $A^{(1)}$
coupling in \p{d10gf} has nothing to do with the PST formulation.
It is a result of the transition from the standard type IIA
supergravity with the RR field strength
$F^{(4)}=dA^{(3)}-H^{(3)}\wedge A^{(1)}$ and the NS--NS two--form
field $B^{(2)}$ to the dual formulation with the six--form gauge
field $B^{(6)}$. The PST techniques has just allowed us to get
this formulation and corresponding gauge and supersymmetry
transformations of fields in a relatively simple way.

In conclusion of this section let us discuss the truncation of our
model to $N=1$, $D=10$ supergravity. To this end in \p{d10st}
and/or in \p{d10gf} we should set
 to zero the gauge fields
$A^{(1)}$ and $A^{(3)}$ together with the left--handed gravitino
and the right--handed dilatino which implies that
\begin{equation}\plabel{psiL}
\psi_{L}=0 \leftrightarrow \psi=\G^{  11}\psi,\qquad \l_{R}=0
\leftrightarrow \l=-\G^{  11}\l.
\end{equation}
After that we arrive at the following action for $N=1$, $D=10$
supergravity with the six--index photon instead of $B^{(2)}$
proposed in \cite{chams}
$$ S=\int_{{\cal M}^{10}}\,\left[-R^{{a}_1{ a}_2}\wedge \Sigma_{{ a}_1{ a}_2}
-{i\over 3!}\bar{\psi}\wedge {\cal D}\psi \G^{{a}_1{a}_2{ a}_3}
\wedge  \Sigma_{{  a}_1{  a}_2{  a}_3}-{i\over
2}\bar\l\G^{  a}{\cal D}\l\wedge \Sigma_{  a}\right]
 $$
$$ +\int_{{\cal M}^{10}}\,\left[{1\over 2}d\phi \wedge \ast
d\phi-C^{(1)}\ast d\phi+{1\over 2}e^{-\phi}dB^{(6)}\wedge \ast
dB^{(6)}+(C^{(3)}-e^{-\phi}\ast C^{(7)}) \wedge
dB^{(6)}\right]+{\cal O}(f^4)\,, $$ where $$ C^{(1)}={i\over
2}\bar{\psi}\l, $$ $$ C^{(3)}={1\over 4}e^{-{1\over
2}\phi}\bar{\psi}\wedge\G^{(1)}\wedge\psi +{1\over 4}e^{-{1\over
2}\phi}\bar{\psi}\wedge\G^{(2)}\l, $$ $$ C^{(7)}={i\over
4}e^{{1\over 2}\phi}\bar{\psi}\wedge\G^{(5)}\wedge\psi+{i\over
4}e^{{1\over 2}\phi}\bar{\psi}\wedge\G^{(6)}\l. $$ As in the whole
paper we have hidden the quartic fermion terms under the ${\cal
O}(f^4)$.

Therefore, one of the dual versions of type IIA supergravity
considered above is an $N=2$ generalization of $N=1$, $D=10$
supergravity by Chamseddine.

\section{Conclusion}

To summarize, we have constructed the duality--symmetric version
of type IIA $D=10$ supergravity which in its final form contains
in addition to the standard type IIA supergravity bosonic fields
also their duals. Although we have not included into consideration
the quartic fermion terms this part of the action remains the same
as that of the standard type IIA supergravity \cite{gp,cw,hn}. We
have analyzed the symmetry structure of this formulation and its
equations of motion, and have established its relation to the
conventional type IIA supergravity
as well as to the doubled field formalism by Cremmer, Julia, L\"u and Pope,
which we lifted off--shell, to the level of the covariant
actions. We have also obtained a new
dual version of type IIA supergravity with the six--form gauge
field instead of the NS--NS two--form which is the $N=2$ extension
of $N=1$ $D=10$ supergravity by Chamseddine.

Another possible truncation of the duality--symmetric action
\p{compl} to $N=1$, $D=10$ would be to keep, upon solving part of
the duality relations, the six--form and the eight--form gauge
field. Remember that the latter is dual to the dilaton. In this
way one gets the dual version of $N=1$ $D=10$ supergravity whose
superfield formulation was considered in \cite{km}.

One can regard the results of this paper as lifting onto the level
of the proper action the on shell constructions of
\cite{cjlp,llps} and \cite{bkorvp}. The coupling of this
duality--symmetric type IIA supergravity to the Dp--branes and to
the NS5--brane can be carried out in a conventional way. Another
advantage of our formulation is that the type IIA action is
written in a form similar to that of type IIB supergravity
\cite{dlt}, which allows one to directly verify the T--duality of
the whole supersymmetric sectors of these theories.

\subsection*{Acknowledgments}
The authors are thankful to Oleg Andreev, Xavier Bekaert,
Gianguido Dall'Agata, Paolo Pasti, Mario Tonin and Mirian Tsulaia
for interest to this work and useful discussions. D.S. would like
to thank Dieter Luest for the hospitality extended to him at the
Institute of Physics of Humboldt University where this work was
finished, and the A. von Humboldt Foundation for financial
support. This work was also partially supported by the Grant N
F7/336-2001 of the Ukrainian State Fund for Fundamental Research,
by the INTAS Research Project N 2000-254 (I.B.,A.N.,D.S.),
BFM2002-03681 grant from the Ministerio de Ciencia y
Tecnolog\'{\i}a de Espa\~na and from EU FEDER funds (I.B.) and by
the European Commission TMR Programme HPRN-CT-2000-00131 and
MURST--COFIN contract 2001-025492 to which D.S. is associated with
the University of Padua.

\newpage

\section*{Appendix A. Notation and conventions}
\def\theequation{A.\arabic{equation}}
\setcounter{equation}0
 In the description of dimensional reduction
(from D=11 to D=10) we use the hat symbol $\hat{}$ to distinguish
the eleven--dimensional quantities (D=11 coordinates, fields,
forms and indices), {\it e.g.} ${X}^{ {\hat m}}= (x^{ {m}}, X^{
\underline {11}})$. We use letters from the middle of the Latin
alphabet for the world indices and from the beginning of the
alphabet for the Lorentz indices. We have underlined the world
index of the compactified coordinate $X^{ \underline {11}}$ to
distinguish it from the corresponding Lorentz index like that of
$\Gamma^{11}$.

We use the mostly minus signature ${\eta}_{{ a}{  b}}= diag
(+---\dots -)$. The antisymmetric $D$--dimensional Levi--Civita
tensor $\e^{ {a_1}\dots {a_D}}$
is defined by
\begin{equation}\plabel{LC}
\e^{ 0 {1}\dots {(D-1)}}=1 \; , \qquad \e_{ 0{1}\dots
{(D-1)}}=(-1)^{D-1} \; ,
\end{equation}
so that
\begin{equation}\plabel{LCc}
\e^{ {a_1}\dots {a_D}} \e_{ {a_1}\dots {a_D}}=(-)^{D-1}D!.
\end{equation}
For an arbitrary $n$--form we have
\begin{equation}\plabel{nform1}
 {F}^{(n)}=  {1\over n!} dx^{ {  m}_n}\wedge\dots \wedge dx^{ {
m}_1} {F}^{(n)}_{ {  m}_1\dots {  m}_n} = {1\over n!}  {E}^{ {
a}_n}\wedge\dots \wedge {E}^{ {  a}_1} {F}^{(n)}_{ {  a}_1\dots {
a}_n}\,,
\end{equation}
and the exterior derivative $d=dx^{ {  m}}
\partial_{ {  m}}$ acts from the right.

The Hodge star operation is defined as follows
\begin{equation}\plabel{FhodgeDA}
( {\ast}  {F}^{(n)})_{ {  a}_1\ldots {  a}_{D-n}}:= { {\a}_n\over
n!} \e_{ {  a}_1\dots {  a}_{D-n} {  b}_1\dots {  b}_n}
 {F}^{(n) {  b}_1\dots {  b}_n} \; ,
\end{equation}
or, equivalently,
\begin{equation}\plabel{EhodgeDA}
 {\ast} ( {E}^{ {  b}_n}\wedge\dots \wedge {E}^{ {
b}_1})= { {\a}_n\over (D-n)!}  {E}^{ {  a}_{D-n}}\wedge\dots \wedge
 {E}^{ {  a}_1}{\e_{ {  a}_1\dots {  a}_{D-n}}}^{
{  b}_1\dots {  b}_n}\; .
\end{equation}
This implies
\begin{equation}\plabel{FhDA}
 {\ast}  {F}^{(n)}  = { {\a}_n\over n! (D-n)!}  {E}^{
{  a}_{D-n}}\wedge\dots \wedge {E}^{ {  a}_1} \e_{ {  a}_1\dots {
a}_{D-n} {  b}_1\dots {  b}_n}  {F}^{(n) {  b}_1\dots { b}_n}\; .
\end{equation}

The coefficients ${ \a_n}$ can be fixed to obey
\begin{equation}\plabel{alphas}
{ \a_n}{ \a_{D-n}}=(-)^{(D-n)n+(D-1)},
\end{equation}
which provides the universal identity
\begin{equation}\plabel{Hodge2A}
{ \ast}{ \ast}=1\; .
\end{equation}
 In odd space--time dimensions all $\alpha_n$ are equal to
one, while in even dimensions we have a freedom in fixing their
values. For instance, in $D=10$ we choose
\begin{equation}\plabel{alphas1}
\a_1=\a_2=\a_3=\a_4=\a_5=\a_7=\a_9=1\; , \qquad \a_6=\a_8=-1\; .
\end{equation}
Note that \p{EhodgeDA} implies
\begin{equation}\plabel{dXhodgeDA}
 {\ast} (dx^{ {  n}_n}\wedge\dots \wedge dx^{ {  n}_1})= {1\over
(D-n)!}{ {\a}_n\over{\sqrt{|g|}}}dx^{ {  m}_{D-n}}\wedge\dots
\wedge dx^{
{ m}_1}{\e_{ {  m}_1\dots {  m}_{D-n}}}^ { {  n}_1\dots { n}_n},
\end{equation}

In our notation
$$
dX^{ {  m}_1}\wedge\dots \wedge dX^{ { m}_D}=
d^D X~ \e^{ {  m}_1\dots {  m}_D},
$$ $$  {E}^{ {  a}_1}\wedge\dots\wedge
 {E}^{ {  a}_D}=d^D X~ \det{  E}~ \e^{ {  a}_1\dots
{  a}_D}.
$$
and
\begin{eqnarray}\plabel{kinA}
&& \int_{{\cal M}^D}\, \ast F^{(n)}\wedge F^{(n)}=
(-)^{D(D-1)\over 2} {1\over n!\a_{D-n}}\int\,d^Dx\,\sqrt{|g|} F_{
{m_1}\dots {m_n}} F^{ {m_1}\dots {m_n}}\; ,
\\ \plabel{con}
&& \int_{{\cal M}^D}\, \omega\wedge\ast\psi= \int_{{\cal M}^D}\,
\psi\wedge\ast\omega\; .
\end{eqnarray}

Note also that for any contravariant vector $V^{  a}$ associated
with the one--form $V^{(1)}= E^{  a}V_{  a}$ and any
$D$--dimensional form $\omega^{(n)}$ one can prove the following
useful identities
\begin{equation}\plabel{id8}
i_V\ast\omega^{(n)}=(-)^{D-n-1}{\a_n\over
\a_{n+1}}\ast(\omega^{(n)}\wedge V^{(1)})\; ,
\end{equation}
\begin{equation}\plabel{id9}
\ast i_V \omega^{(n)}=(-)^{D-n}{\a_{n-1}\over
\a_{n}} *\omega^{(n)}\wedge V^{(1)}\; ,
\end{equation}
where the contraction is defined by
\begin{eqnarray}\plabel{contr}
i_V \omega^{(n)} &\equiv& V^a i_a\om^{(n)}\equiv V^m i_m\om^{(n)} \; ,
\qquad \nonumber \\
i_a\om^{(n)}& =& {1\over (n-1)!} E^{a_{n-1}}\wedge \ldots \wedge  E^{a_{1}}
\omega_{a a_{1}\ldots a_{n-1}}\; ,
\qquad
\nonumber \\
i_m\om^{(n)}&=& {1\over (n-1)!} dx^{m_{n-1}}\wedge \ldots \wedge  dx^{m_{1}}
\omega_{m m_{1}\ldots m_{n-1}}\; . \qquad
\end{eqnarray}
In particular, in D=10 we have
\begin{equation}\plabel{idD10}
D=10\, : \qquad *(*H^{(7)}\wedge A^{(1)}) = i_A H^{(7)}\;  , \quad
*(*F^{(4)}\wedge A^{(1)}) = i_A F^{(4)}\; .
\end{equation}
One can also check that, in any $D$,
\begin{eqnarray}\plabel{id10}
i_V \left(\om^{(n)} -  {i_V \om^{(n)} \wedge V^{(1)}\over V^2}\right)=0\; ,
\\ \plabel{id11}
*i_V(\om^{(n)}\wedge V^{(1)}) = i_V*\om^{(n)}\, \wedge V^{(1)}\; .
\end{eqnarray}

\subsection*{Gauge field strengths of duality--symmetric type IIA
supergravity}

In addition to the gravitational field, the conventional bosonic
fields are $$\phi(x), \quad A^{(1)}(x), \quad B^{(2)}(x), \quad
A^{(3)}(x)$$ and their duals are, respectively $$ A^{(8)}(x),
\quad A^{(7)}(x), \quad B^{(6)}(x), \quad A^{(5)}(x). $$ The field
strengths of the dual pairs are $$
\!\!\!\!\!\!\!\!\!\!F^{(1)}=d\phi\,, \quad\quad\quad\quad \quad
\quad \quad F^{(9)}=dA^{(8)}-{3\over 4}F^{(8)}\wedge
A^{(1)}+{1\over 2}B^{(2)}\wedge dB^{(6)}-{1\over 4}F^{(6)}\wedge
A^{(3)}, $$
$$
\!\!\!\!\!\!\!\!\!\!\!\!\!\!\!\!\!\!\!\!\!\!\!\!\!\!\!\!\!\!\!\!\!\!\!\!\!\!\!\!\!\!
F^{(2)}=dA^{(1)}\; , \quad\quad\quad \quad \qquad
F^{(8)}=dA^{(7)}+F^{(6)} \wedge B^{(2)}+ B^{(2)} \wedge B^{(2)}
\wedge dA^{(3)}\,, $$
\begin{equation}\plabel{gfs}
\!\!\!\!\!\!\!\!\!\!\!\!\!\!\!\!\!\!\!\!\!\!\!\!\!\!\!\!\!\!\!\!\!\!\!\!\!\!\!\!
\!\!\!\!\!\!\!\!\!\!\!\!\!\!\!\!\!\!\!\!
H^{(3)}=dB^{(2)},\quad\quad\quad\quad\quad\quad H^{(7)} =
dB^{(6)}+A^{(3)} \wedge dA^{(3)}-F^{(6)} \wedge A^{(1)}\,,
\end{equation}
$$
\!\!\!\!\!\!\!\!\!\!\!\!\!\!\!\!\!\!\!\!\!\!\!\!\!\!\!\!\!\!\!\!\!\!\!\!
\!\!\!\!\!\!\!\!\!\!\!\!\!\!\!\!\!\!\!\!\!\!\!\!
F^{(4)}=dA^{(3)}-H^{(3)}\wedge A^{(1)}, \quad F^{(6)} =
dA^{(5)}+A^{(3)}\wedge H^{(3)}-B^{(2)}\wedge dA^{(3)}\,.
$$
\subsection*{Gauge field strengths in Romans's supergravity}

\begin{eqnarray}\plabel{gfsr}
& F_m^{(1)}=F^{(1)}=
d\phi\; ,
\qquad & F^{(9)}_m=F^{(9)}-{5\over 4}mA^{(9)}-{1\over 2}m
A^{(7)}\wedge B^{(2)} +{1\over 16}mA^{(1)}\wedge (B^{(2)})^{\wedge
4}\; ,
\nonumber \\
& F^{(2)}_m=F^{(2)}-mB^{(2)}\; ,
\qquad  & F^{(8)}_m=F^{(8)}-{1\over 12}m(B^{(2)})^{\wedge 4}\,,
\nonumber \\
& H^{(3)}_m=H^{(3)}=
dB^{(2)}\; ,
\qquad
& H^{(7)}_m
=H^{(7)}-mA^{(7)} + {1\over 3}mA^{(1)}\wedge (B^{(2)})^{\wedge
3}\,,
\nonumber \\
& F^{(4)}_m=F^{(4)}+{1\over 2}m(B^{(2)})^{\wedge 2}\,, \quad
&F^{(6)}_m =F^{(6)}- {1\over 3}m(B^{(2)})^{\wedge 3}\,,
\end{eqnarray}
where $m$ is the mass parameter and the field strengths $F^{(n)}$
and $H^{(n)}$ are defined in eq. \p{gfs}.

When the mass parameter $m$ is promoted to the field $F^{(0)}(x)$,
one should replace $m$ with  $F^{(0)}(x)$ in the definition of the
field strengths \p{gfsr} and introduce  the field strength
$F^{(10)}_m$ dual to $F^{(0)}(x)$
$$
F^{(10)}_m=d{A}^{(9)}+B^{(2)}\wedge F^{(8)}-{1\over
2}(B^{(2)})^{\wedge 2}\wedge F^{(6)}-{1\over 3}(B^{(2)})^{\wedge
3}\wedge dA^{(3)}-{1\over 60}F^{(0)}(B^{(2)})^{\wedge 5}\, .
$$

\section*{Appendix B. Dimensional reduction of Einstein-Hilbert term}
\def\theequation{B.\arabic{equation}}
\setcounter{equation}0

To dimensionally reduce the action for $D=11$ supergravity, let us
begin with a general representation of the $D$-dimensional line
element
\begin{equation}\plabel{inta}
d{\hat s}^2\equiv {\hat g}^{(D)}_{ {\hat m} {\hat n}} dX^{ {\hat
m}}\otimes dX^{ {\hat n}}= e^{2 \a \phi(x)}g^{(D-1)}_{ {m} {n}}dx^{
{m}}\otimes dx^{ {n}}-e^{2 \b \phi}(dX^{\underline{11}}+A^{(1)})
\otimes (dX^{\underline {11}}+A^{(1)}),
\end{equation}
where, as in $D=11$, we have defined the compactified coordinate
by $X^{\underline{11}}$.
 This choice corresponds to the following splitting of the
$D$--dimensional vielbein one--form
\begin{equation}\plabel{viela}
{\hat E}^{  a}=e^{\a \phi (x)}dx^{  m}E_{  m}^ {~  a}(x) \, ,\qquad {\hat
E}^{ {11}}=e^{\b \phi(x)}(dX^{\underline{11}}+A^{(1)}).
\end{equation}
The torsion two--form is
\begin{equation}\plabel{torsa}
{\hat T}^ { {\hat a}}:=
d{\hat E}^{ {\hat a}}-{\hat E}^{ {\hat b}}\wedge {\hat \omega}_{
{\hat b}}^{~ {\hat a}}={i\over 4}\hat{\bar
\Psi}\G^{ {\hat a}}\wedge\hat{\Psi}.
\end{equation}

Splitting the indices and using the following ansatz for the
 gravitino field
\begin{equation}\plabel{gransB}
\hat{\Psi}=e^{-{7\a+\b \over 2}\phi}(\psi+\a\G^{(1)}\G^{11}\l) +\b
e^{{7\a+3\b \over 2}\phi}\l(dX^{\underline {11}}+A^{(1)})
\end{equation}
we derive the components of the connection one--form defining the
curvature two--form $\hat{R}^{ {\hat a} {\hat b}}=d\hat{\omega}^{
{\hat a} {\hat b}} -{\hat{\omega}^{ {\hat a}}}_{~ {\hat c}}\wedge
\hat{\omega}^{ {\hat c} {\hat b}}$ $$ {\hat \omega}_{ {bc}}^{~~ a}
=e^{-\a\phi}(\omega_{ {bc}}^{~~ a}+2\a
\partial_{[ {b}}\phi \d_{ {c}]}^{~  a})+{i\over 4}e^{-(\b+9\a)\phi}
(\bar{\psi}_b\G^a\psi_c-2\a\bar{\psi}_b(\G_c^{~a}\G^{11})\l
-2\a\d^a_{[b}\bar{\psi}_{c]}\G^{11}\l-\a^2\bar{\l}\G_{bc}^{~~a}\l),
$$ $${\hat \omega}_{ {11} {b}}^{~~ {11}}=-\b
e^{-\a\phi}\partial_{  b}\phi-{i\over 2}\b
e^{-\a\phi}\bar{\psi}_b\G^{11}\l,$$ $$ {\hat \omega}_{ {a}
{b}}^{~~ {11}}={1\over 2} e^{(\b-2\a)\phi}F^{(2)}_{ {ab}}+{i\over
4}e^{-(\b+9\a)\phi}(\bar{\psi}_a\G^{11}\psi_b+2\a\bar{\psi}_a\G_b\l
+\a^2\bar{\l}(\G_{ab}\G^{11})\l),$$
\begin{equation}\plabel{conna}
\hat{\omega}_{ {11} {a} {b}} ={\hat{\omega}_{ {a} {b}}}^{~~
{11}}-{i\over
2}e^{-\a\phi}(\b\bar{\psi}_a\G_b\l+\a\b\bar{\l}(\G_{ab}\G^{11})\l),
\end{equation}
where $F_{ {a} {b}}^{(2)}$ is the field strength of the KK vector
field $A^{(1)}$.

Since, up to a surface term, the torsion enters the
Einstein--Hilbert action only in quadratic
combinations (see e.g.
\cite{hn}) one can neglect it (and, hence, the fermion inputs
into the spin connection) in the quadratic fermion
approximation.

By use of the Palatini identity \cite{bho} (the numerical
coefficient $\Delta$ is equal to zero for D=11)
$$
\int_{{\cal M}^D}
{1\over (D-2)!} e^{\Delta\phi} {\hat R}^{ {\hat{a}_1}
{\hat{a}_2}}(\hat{\omega})\wedge \hat{E}^{ {\hat a}_3}
\wedge\dots \wedge \hat{E}^{ {\hat a}_D} \e_{ {\hat{a}_1}\dots
{\hat{a}_D}} \equiv (-)^{D} \int\, d^D x \,
\sqrt{-\hat{g}}e^{\Delta \phi} {\hat R}
$$
\begin{equation}\plabel{Pala}
=(-)^{D}\int\,\,d^D x~ \det{\hat E}e^{\Delta\phi} [\hat{\omega}_{
{\hat b}}^{~ {\hat b} {\hat a}} \hat{\omega}_{ {\hat c}~~ {\hat
a}}^{~ {\hat c}} +\hat{\omega}_{ {\hat a}}^{~ {\hat b} {\hat c}}
\hat{\omega}_{ {\hat b} {\hat c}} ^{~~ {\hat a}}+2\Delta
\hat{\omega}_{ {\hat b}}^{~ {\hat b} {\hat a}}
\partial_{ {\hat a}}\phi],
\end{equation}
 and
the expression for the connection coefficients \p{conna}, after
some algebra we arrive at the following intermediate form of the
dimensionally reduced Einstein--Hilbert term
$$
\int_{{\cal M}^{D}}\,{\cal L}^{(D)}_{EH}= (-)^{D}\int\,
dX^{\underline {11}} \int\,d^{D-1}x~e^{((D-3)\a+\b)\phi}
\sqrt{|g|}\{ {\omega}_{ {b}}^{~ {b} {a}} {\omega}_{ {c}~~ {a}}^{~
{c}}+ {\omega}_{ {a}}^{~ {b} {c}} {\omega}_{ {b} {c}} ^{~~ {a}} $$
$$ +2(\a (D-3)+\b) {\omega}_{ {b}}^{~ {b} {a}}
\partial_{  a}\phi
+(\a (D-2)+\b)^2 (\partial \phi)^2-\a^2(D-2)(\partial \phi)^2
-\b^2(\partial \phi)^2 $$
\begin{equation}\plabel{EHa}
+{1\over 4}e^{2(\b-\a)\phi}F^{(2)}_{ {ab}}F^{(2)  {ab}} \}.
\end{equation}
Applying the Palatini identity backwards,  we finally obtain
$$
\int_{{\cal M}^D} {1\over (D-2)!} e^{\Delta\phi} {\hat R}^{
{\hat{a}_1}  {\hat{a}_2}}(\hat{\omega})\wedge \hat{E}^{ {\hat
a}_3}(d) \wedge\dots \wedge \hat{E}^{ {\hat a}_D}(d) \e_{
{\hat{a}_1}\dots {\hat{a}_D}}
$$
$$
=(-)^{D}
\int\,d^{(D-1)}x~ e^{((D-3)\a+\b)\phi} \sqrt{|g|}\{R+(\a
(D-2)+\b)^2 (\partial \phi)^2 $$
\begin{equation}\plabel{EHRa}
-\a^2(D-2)(\partial \phi)^2 -\b^2(\partial \phi)^2 +{1\over
4}e^{2(\b-\a)\phi}F^{(2)}_{ {ab}}F^{(2)  {ab}} \}.
\end{equation}
Note that from the beginning we set the gravitational coupling
constant and the compactification radius $r=\int\,d X^{ {11}}$ to
one.

To get both the $D$--dimensional and the $(D-1)$--dimensional
actions written in the Einstein frame, where the Einstein--Hilbert
term does not include an input from dilaton(s), one should assume
(see \cite{lpss})
\begin{equation}\plabel{ABa}
\a^2={1\over 2(D-2)(D-3)};\qquad \b=-(D-3)\a \; .
\end{equation}
Indeed, in this case one obtains from \p{EHRa}
\begin{equation}\plabel{EHR1a}
\int\, d^D x \sqrt{-{\hat g}}{\hat R}=
\int\, d^{(D-1)} x~ \sqrt{|g|} [R-{1\over 2}(\partial
\phi)^2+{1\over 4}e^{-2(D-2)\a\phi}F^{(2)2}].
\end{equation}
Recall that there is no dilaton in $D=11$ supergravity multiplet
and hence $\Delta_{D=11}=0$. With the choice of $\a=+1/12$ one
gets also the Einstein frame form for the $10$--dimensional
action,
\begin{equation}\plabel{EHR2a}
\int\, d^{11} x \sqrt{-{\hat g}}{\hat R}=
\int\, d^{10} x~ \sqrt{|g|} [R-{1\over 2}(\partial \phi)^2+{1\over
4}e^{-{3\over 2}\phi}F^{(2)2}] \; .
\end{equation}

\section*{Appendix C. Dimensional reduction of antisymmetric tensor fields}
\def\theequation{C.\arabic{equation}}
\setcounter{equation}0

Under dimensional reduction from $D$ to $D-1$ the $n$-form
potential $\hat{A}^{(n)}$ decomposes as
\begin{equation}\plabel{Cna}
\hat{A}^{(n)}=A^{(n)}-A^{(n-1)}\wedge dX^{\underline {11}},
\end{equation}
where $X^{\underline {11}}$ is the compactified dimension, and the
field strength $\hat{F}^{(n+1)}=d\hat{A}^{(n)}$ is
\begin{equation}\plabel{Fna}
\hat{F}^{(n+1)}=dA^{(n)}+dA^{(n-1)}\wedge dX^{\underline {11}}=
F^{(n+1)}+F^{(n)}\wedge (dX^{\underline {11}}+A).
\end{equation}

By use of Appendix A and the representation for the interval
\p{inta} one gets the following expression for the dual field
strength
\begin{equation}\plabel{Fn*Da}
{\hat \ast}{\hat F}^{(n)}=(-)^{D}{\hat{\a}_n\over \a_n}
e^{(D-2n-1)\a\phi+\b\phi} \ast F^{(n)}\wedge (dX^{\underline
{11}}+A) +(-)^{n-1}{\hat{\a}_{n} \over \a_{n-1}}
e^{(D-2n+1)\a\phi-\b\phi}\ast F^{(n-1)}\, .
\end{equation}
In the Einstein frame \p{Fn*Da} is
\begin{equation}\plabel{Fn*D1a}
{\hat \ast}{\hat F}^{(n)}=(-)^{D}{\hat{\a}_n\over \a_n}
e^{-2(n-1)\a\phi} \ast F^{(n)}\wedge (dX^{\underline {11}}+A)
+(-)^{n-1}{\hat{\a}_{n} \over \a_{n-1}} e^{2(D-n-1)\a\phi}\ast
F^{(n-1)}.
\end{equation}
Taking into account this relation and separating the part
containing $dX^{ {11}}$ we get the gauge field kinetic terms in
the form $$ \int_{{\cal M}^D}\, {\hat F}^{(n)}\wedge {\hat \ast}
{\hat F}^{(n)} =(-)^{D}[{\hat{\a}_n \over \a_n} \int_{{\cal
M}^{(D-1)}}\, e^{(D-2n-1)\a\phi+\b\phi} F^{(n)} \wedge \ast
F^{(n)} $$
\begin{equation}\plabel{kinFEa}
-{\hat{\a}_{n} \over \a_{n-1}} \int_{{\cal M}^{(D-1)}}\,
e^{(D-2n+1)\a\phi-\b\phi} F^{(n-1)} \wedge \ast F^{(n-1)}] \cdot
\int_{{\cal M}^1}dX^{ \underline{11}}\, .
\end{equation}
In the Einstein frame, {\it i.e.} with  \p{ABa}, eq. \p{kinFEa}
becomes $$ \int_{{\cal M}^{D}}\,{\hat
F}^{(n)}\wedge{\hat\ast}{\hat F}^{(n)} =(-)^{D}[{\hat{\a}_n \over
\a_n} \int_{{\cal M}^{D-1}}\, e^{-2(n-1)\a\phi} F^{(n)}\wedge\ast
F^{(n)} $$
\begin{equation}\plabel{kinFE1a}
-{\hat{\a}_{n} \over \a_{n-1}} \int_{{\cal M}^{(D-1)}}\,
e^{2(D-n-1)\a\phi}F^{(n-1)}\wedge\ast F^{(n-1)}] \cdot \int_{{\cal
M}^{1}}\, dX^{\underline{11}}.
\end{equation}

\section*{Appendix D. Useful identities}
\def\theequation{D.\arabic{equation}}
\setcounter{equation}0

The variational problem for the duality--symmetric part of the
supergravity action may be simplified by considering some special
identities which hold for any values of space-time dimension $D$
and of the rank $n$ of differential forms. To be precise, let us
consider the following variation
\begin{equation}\plabel{var}
\d (v\wedge F^{(n)}\wedge i_v {\cal F}^{(D-n)})
\end{equation}
with the one--form $v$ defined in \p{v1f} and ${\cal
F}^{(D-n)}=F^{(D-n)}- \beta \ast F^{(n)}$. We have
$$
\d (v\wedge F^{(n)}\wedge i_v {\cal
F}^{(D-n)})=\d v\wedge F^{(n)}\wedge i_v {\cal F}^{(D-n)}+v\wedge
\d F^{(n)}\wedge i_v {\cal F}^{(D-n)}
$$
\begin{equation}\plabel{var1}
+v\wedge F^{(n)}\wedge i_{\d v} {\cal F}^{(D-n)} +v\wedge
F^{(n)}\wedge i_v \d {\cal F}^{(D-n)}
\end{equation}
Consider now the first term of the last line of \p{var1}
\begin{equation}\plabel{1B}
v\wedge F^{(n)}\wedge i_{\d v} {\cal F}^{(D-n)}= v\wedge
F^{(n)}\wedge i_{\d v} F^{(D-n)}- \beta v\wedge F^{(n)}\wedge
i_{\d v} \ast F^{(n)}.
\end{equation}
Using the conventions and identities
listed in Appendix A, by straightforward
calculations we get
\begin{equation}\plabel{id1}
v\wedge F^{(n)}\wedge i_{\d v}
F^{(D-n)}=- (-)^{D(n+1)} \d v\wedge v \wedge i_v \ast
F^{(n)}\wedge i_v \ast F^{(D-n)}
\end{equation}
and
\begin{equation}\plabel{id2}
v\wedge F^{(n)}\wedge i_{\d v} \ast F^{(n)}=- (-)^{D(n+1)}
\d v\wedge v \wedge i_v \ast F^{(n)}\wedge i_v F^{(n)}.
\end{equation}

The last term in \p{var1} is
\begin{equation}\plabel{2B}
v\wedge F^{(n)}\wedge i_v \d {\cal F}^{(D-n)}= v\wedge
F^{(n)}\wedge i_v \d F^{(D-n)}- \beta v\wedge F^{(n)}\wedge
i_v \d \ast F^{(n)} \; .
\end{equation}
The first term in the {\it r.h.s} can be simplified using the
identity
$$ 0=  i_v(v\wedge F^{(n)}\wedge \d F^{(D-n)})=
v\wedge F^{(n)}\wedge i_v \d F^{(D-n)}
$$
\begin{equation}\plabel{id3}
+(-)^{D-n} v\wedge i_v F^{(n)}\wedge \d
F^{(D-n)} - (-)^{D}F^{(n)}\wedge \d F^{(D-n)}\; ,
\end{equation}
(remember that $i_v v=-1$), namely
\begin{equation}\plabel{id4}
v\wedge F^{(n)}\wedge i_v \d F^{(D-n)}= (-)^{D}F^{(n)}\wedge \d
F^{(D-n)}- (-)^{D-n} v\wedge i_v F^{(n)}\wedge \d F^{(D-n)}\; .
\end{equation}
In view of \p{id3}, assuming that $\delta g_{mn}=0$, one gets
for the second term of \p{2B}
\begin{eqnarray}\plabel{nid5}
v\wedge F^{(n)}\wedge i_v \d \ast F^{(n)}= -(-)^{D-n} i_v (v\wedge
F^{(n)}) \wedge \d \ast F^{(n)} = \nonumber \\ = -(-)^{D-n} \d
F^{(n)}\wedge *i_v (v\wedge F^{(n)}) =
 -(-)^{D(n+1)+n} *i_v (F^{(n)}\wedge v) \wedge  \d  F^{(n)}\; .
\end{eqnarray}
Then, using the identity \p{id11},
$ *i_v (F^{(n)}\wedge v)= i_v *F^{(n)}\, \wedge v$, one finally obtains
\begin{eqnarray}\plabel{nid6}
v\wedge F^{(n)}\wedge i_v \d \ast F^{(n)}= (-)^{Dn}
v\wedge i_v \ast F^{(n)} \wedge \d F^{(n)} \; ,
\end{eqnarray}
which completes the reduction of the variational problem for a PST
action to the standard one.

\section*{Appendix E. Gamma--matrix conventions}
\def\theequation{E.\arabic{equation}}
\setcounter{equation}0

We use the following conventions for the Gamma matrices (see,
e.g., \cite{nof})
$$ \{\G^a,\G^b\}=2\eta^{ab},
$$
$$ \G^{a_j\dots a_2 a_1}\G_{b_1 b_2\dots
b_k}=\sum^{min(j,k)}_{l=0}\, l!
\left(\begin{array}{c}j\\l\end{array}\right)
\left(\begin{array}{c}k\\l\end{array}\right)
\d^{[a_1}_{[b_1}\dots\d^{a_l}_{b_l}{\G^{a_j\dots
a_{l+1}]}}_{~b_{l+1}\dots b_k]},\qquad
\left(\begin{array}{c}j\\l\end{array}\right) ={j!\over l!(j-l)!},
$$
$$
\G^{(n)}={1\over n!}E^{a_n}\wedge\dots E^{a_1}\G_{a_1\dots
a_n}
$$
in any space--time dimension.

In $D=11$ we define
$$
\G_{\hat{  a}_1\dots\hat{ a}_n}={i\over
(11-n)!}(-)^{{(11-n)(10-n)\over 2}}\e_{\hat{ a}_1\dots\hat{
a}_n\hat{  b}_1\dots \hat{  b}_{11-n}}\G^{\hat{ b}_1\dots \hat{
b}_{11-n}}.
$$
In ten space--time dimensions we also have
$$
\G^{11}=-i\G^{0}\G^{1}\dots\G^{9},\qquad \{\G^{ a},\G^{
11}\}=0,\qquad (\G^{  11})^2=-1,
$$
and
$$
\bar{\psi}_1\G^{{  a}_1\dots{
a}_n}\psi_2= (-)^n\bar{\psi}_2\G^{{  a}_n\dots{ a}_1}\psi_1,
$$
$$
\bar{\psi}_1\G^{{  a}_1\dots{ a}_n}\G^{  11}\psi_2=
-\bar{\psi}_2\G^{{ a}_n\dots{ a}_1}\G^{ 11}\psi_1
$$
for two Majorana spinors.

\end{document}